\documentclass[12pt]{article}
\usepackage{epsfig}
\usepackage{hyperref}
\usepackage{amssymb}
\usepackage{colordvi}

\newcommand{\be}{\begin{eqnarray}}
\newcommand{\ee}{\end{eqnarray}}
\newcommand{\bite}{\begin{itemize}}
\newcommand{\eite}{\end{itemize}}

\newcommand{\csw}{c_{SW}}

\newcommand{\itep}
{~\vspace{-1.2cm}
\begin{flushright}
{\normalsize DESY 13-046} \\
{\normalsize Edinburgh 2013/03 } \\
{\normalsize Liverpool LTH 972}
\end{flushright}

\vspace{0.0cm}}
\begin{document}

\renewcommand{\thefootnote}{\alph{footnote}}
\setcounter{footnote}{0}

\begin{center}
\itep

{\Large\bf
Perturbatively improving RI-MOM renormalization constants
}
\vspace*{0.4cm}

{\large
M.~Constantinou$^1$,  M.~Costa$^1$, M.~G\"ockeler$^2$, R.~Horsley$^3$,
H.~Panagopoulos$^1$,
H.~Perlt$^{4}$,  P.~E.~L.~Rakow$^5$, G.~Schierholz$^{6}$  and A.~Schiller$^4$
}

\vspace*{0.4cm}

{\sl
$^1$ Department of Physics, University of Cyprus, 
     P.O.Box 20537, Nicosia CY-1678, Cyprus \\
$^2$ Institut f\"ur Theoretische Physik, Universit\"at
Regensburg,  93040 Regensburg, Germany \\
$^3$ School of Physics, University of Edinburgh, 
     Edinburgh EH9 3JZ, UK \\
$^4$ Institut f\"ur Theoretische Physik, Universit\"at
Leipzig,  04103 Leipzig, Germany \\
$^5$ Theoretical Physics Division, Department of Mathematical Sciences,
\\ University of Liverpool,  Liverpool L69 3BX, UK \\
$^6$ Deutsches Elektronen-Synchrotron DESY, 
     22603 Hamburg, Germany
}

\vspace*{0.4cm}

\end{center}

\begin{abstract}
The determination of renormalization factors is of crucial importance
in lattice QCD. 
They relate the observables obtained on the lattice to their 
measured counterparts in the continuum in a suitable renormalization scheme. 
Therefore, they have to be computed as precisely as possible. A widely used
approach is the nonperturbative Rome-Southampton method. It requires,
however, a careful treatment of lattice artifacts. 
In this paper we investigate a method to suppress these artifacts by subtracting one-loop
contributions to renormalization factors calculated in lattice perturbation 
theory. We compare results obtained from a complete one-loop subtraction
with those calculated for a 
subtraction of contributions proportional to the square of
the lattice spacing.
\end{abstract}

\section{Introduction}
\label{sec:Intro}

Renormalization factors in lattice Quantum Chromodynamics (QCD) relate 
observables computed on finite lattices to their
continuum counterparts in specific renormalization schemes. Therefore,
their determination should be as precise as possible in order to allow for
a reliable comparison with experimental results. One approach is based
on lattice perturbation theory~\cite{Capitani:2002mp}. 
However, it suffers from its intrinsic
complexity, slow convergence and the impossibility to handle mixing
with lower-dimensional operators. Therefore, nonperturbative methods
have been developed and applied. Among them the so-called
Rome-Southampton method~\cite{Martinelli:1994ty} (utilizing the RI-MOM scheme) is widely used because
of its simple implementation. It requires, however, gauge fixing.

Like (almost) all quantities evaluated in lattice QCD also renormalization
factors suffer from discretization effects. One can attempt to cope
with these lattice artifacts by extrapolating the nonperturbative 
scale dependence to the continuum (see Ref.~\cite{Arthur:2010ht})
or one can try to suppress them by a subtraction procedure based
on perturbation theory. Here we shall deal with the latter approach.

In a recent paper of the QCDSF/UKQCD collaboration~\cite{Gockeler:2010yr}  a 
comprehensive discussion and comparison of perturbative and 
nonperturbative renormalization have been given. Particular emphasis was placed on 
the perturbative subtraction of the unavoidable lattice artifacts.
For simple operators this can be done in one-loop
order completely by computing 
the corresponding diagrams for finite lattice spacing numerically.
While being very effective this procedure is rather involved and
not suited as a general method for more complex operators, especially
for operators with more than one covariant derivative,
and complicated lattice actions.
An alternative approach can be based on the subtraction of
one-loop terms of order $a^2$ with $a$ 
being the lattice spacing. The computation of those
terms has been developed by the authors of Ref.~\cite{Constantinou:2009tr}
and applied to various operators for different actions.
In this paper we use some of those results for the analysis of Monte Carlo
data for renormalization coefficients.

We study the flavor-nonsinglet quark-antiquark operators given in Table~\ref{OpTab}. 
 \begin{table}[!htb] 
   \begin{center}
 \begin{tabular}{|c|c|c|l|}\hline
   Operator  & Notation & Repre- & Operator basis \\ 
   (multiplet)  &   {}  &  sentation & {} \\ \hline 
   $\bar{u}\,d$  & $\mathcal{O}^{S}$        & $\tau_1^{(1)}$ &   $\mathcal{O}^{S}$    \\
   $\bar{u}\,\gamma_\mu\,d$  & $\mathcal{O}_\mu^{V}$        & $\tau_1^{(4)}$ &   $\mathcal{O}_1^{V}, \mathcal{O}_2^{V}, \mathcal{O}_3^{V}, \mathcal{O}_4^{V}$    \\
   $\bar{u}\,\gamma_\mu\gamma_5\,d$  & $\mathcal{O}_\mu^{A}$        & $\tau_4^{(4)}$ &   $\mathcal{O}_1^{A}, \mathcal{O}_2^{A}, \mathcal{O}_3^{A}, \mathcal{O}_4^{A}$    \\
   $\bar{u}\,\sigma_{\mu\nu}\,d$  & $\mathcal{O}_{\mu\nu}^{T}$        & $\tau_1^{(6)}$ &   $\mathcal{O}_{12}^{T}, \mathcal{O}_{13}^{T}, \mathcal{O}_{14}^{T}, \mathcal{O}_{23}^{T}, \mathcal{O}_{24}^{T}, \mathcal{O}_{34}^{T}$    \\ 
   $\bar{u}\,\gamma_\mu \stackrel{\leftrightarrow}{D_\nu}\,d$  & $\mathcal{O}_{\mu\nu}\to\mathcal{O}^{v_{2,a}}$        & $\tau_3^{(6)}$ &   $\mathcal{O}_{\{12\}}, \mathcal{O}_{\{13\}}, \mathcal{O}_{\{14\}}, \mathcal{O}_{\{23\}}, \mathcal{O}_{\{24\}}, \mathcal{O}_{\{34\}}$    \\
   $\bar{u}\,\gamma_\mu \stackrel{\leftrightarrow}{D_\nu}\,d$  & $\mathcal{O}_{\mu\nu}\to\mathcal{O}^{v_{2,b}}$        & $\tau_1^{(3)}$ &   $1/2(\mathcal{O}_{11}+ \mathcal{O}_{22}- \mathcal{O}_{33}- \mathcal{O}_{44})$,    \\ 
    & &  &   $1/\sqrt{2}(\mathcal{O}_{33}- \mathcal{O}_{44}), 1/\sqrt{2}(\mathcal{O}_{11}-\mathcal{O}_{22})$    \\ \hline
\end{tabular} 
   \end{center}
 \caption{Operators and their representations as investigated in the 
present paper. The symbol
$\{...\}$ means total symmetrization. A detailed group theoretical
discussion is given in~\cite{Gockeler:1996mu}.}
 \label{OpTab}
 \end{table}
The corresponding renormalization factors have been measured 
(and chirally extrapolated) at $\beta=5.20, 5.25, 5.29$ and  $5.40$
using $N_f=2$ clover improved Wilson fermions with plaquette gauge 
action~\cite{Gockeler:2010yr}.
All results are computed in Landau gauge. The clover parameter $c_{SW}$
used in the perturbative calculation discussed below is set to its lowest order value $c_{SW}=1$.

\section{Renormalization group invariant operators}
\label{sec:RGI}

We  define the renormalization constant $Z$ of an operator $\mathcal O$ 
from its amputated Green function (or vertex function) $\Gamma (p)$,
where $p$ is the external momentum and
the operator is taken at vanishing momentum. 
The corresponding
renormalized vertex function and the Born term (with all lattice 
artifacts included) are denoted by
$\Gamma_R (p)$ and $\Gamma^{\rm {Born}}(p)$, respectively.
If there is no mixing, $Z$ can then be obtained by imposing the condition
\begin{equation}
\frac{1}{12}\, {\rm tr} \left[\Gamma_R(p)\,\Gamma^{\rm Born}(p)^{-1}\right] = 1
\label{ZDet1}
\end{equation}
for vanishing quark mass at $p^2 = \mu^2$, where $\mu$ is the 
renormalization scale. The $Z$ factor relates the renormalized and 
the unrenormalized vertex function through
\begin{equation}
\Gamma_R(p) = Z_q^{-1}\,Z \, \Gamma(p)\,,
\end{equation}
with $Z_q$ being the quark field renormalization constant determined by
\begin{equation}
Z_q(p)= \frac{{\rm tr}\left[-{\rm i} \sum_\lambda \gamma_\lambda \sin (a p_\lambda) \, 
a S^{-1}(p)\right]}{12\sum_\nu \sin^2(ap_\nu)}
\label{defzq}
\end{equation}
in the chiral limit again at $p^2=\mu^2$. Condition (\ref{ZDet1}) together with
(\ref{defzq}) defines the ${\rm RI}^\prime$-MOM renormalization scheme.
Here $S^{-1}$ is the inverse quark propagator.
Using (\ref{ZDet1}) we compute $Z$ from
\begin{equation}
Z_q^{-1}\,Z\, \frac{1}{12}\,{\rm tr}\left[\Gamma(p)\,\Gamma^{\rm Born}(p)^{-1}\right] = 1\,.
\label{ZDet2}
\end{equation}
For operators transforming as singlets under the hypercubic group $H(4)$,
such as $\mathcal O^{S}$, $Z$ can depend on the components of $p$ only through
$H(4)$ invariants.

For operators belonging to an $H(4)$ multiplet of dimension greater than 1
the condition (\ref{ZDet1}) violates $H(4)$ covariance and would in general 
lead to  different $Z$ factors for each member of the multiplet. In 
Ref.~\cite{Gockeler:2010yr} 
an averaging  procedure has been proposed to calculate one common $Z$ factor for every multiplet. 
Labeling the chosen operator basis by $i=1,2,\ldots,d$ the common 
$Z$ was calculated from
\begin{equation}
  Z_q^{-1}\,Z \,\frac{1}{d} \sum_{i=1}^d\frac{1}{12}\,{\rm tr}\left[\Gamma_i(p)\Gamma_i^{\rm Born}(p)^{-1}\right]=1\,.
\label{ZRinvOld}
\end{equation}
This condition leads to an $H(4)$-invariant $Z$ for the operators
without derivatives in Table~\ref{OpTab}. However, in general this is not the case.

It is not difficult to devise a renormalization condition that
respects the hypercubic symmetry.
Choosing a basis of operators (again labeled by $i$), transforming 
according to a unitary irreducible representation of $H(4)$, the relation
\begin{equation}
   Z_q^{-1}\,Z\,\frac{\sum_{i=1}^d{\rm tr}\left[\Gamma_i(p)\Gamma_i^{\rm Born}(p)^\dagger\right]}
                     {\sum_{j=1}^d{\rm tr}\left[\Gamma_j^{\rm Born}(p)\Gamma_j^{\rm Born}(p)^\dagger\right]}=1\,                 
\label{ZRinv}
\end{equation}
defines a $Z$ factor which is invariant under $H(4)$, provided that the
quark field renormalization factor is also $H(4)$ invariant. 
The derivation of renormalization condition (\ref{ZRinv}) is given in the Appendix.
For the operators without
derivatives the definitions (\ref{ZRinv}) and (\ref{ZRinvOld}) are
equivalent. 
For the considered operators with one 
derivative the resulting differences turn out to be negligible.
In the following the $Z$ factors will be determined 
from (\ref{ZRinv}) using the operator bases given in Table~\ref{OpTab}. 
This is our version of the ${\rm RI}^\prime$-MOM scheme. 

We define a so-called RGI (renormalization group invariant) operator, which is independent
of scale $M$ and scheme $\mathcal{S}$, by~\cite{Gockeler:2010yr}
\begin{equation}
\mathcal{O}^{\rm RGI} = \Delta Z^{\mathcal{S}}(M)\, \mathcal{O}^{\mathcal{S}}(M) = Z^{\rm RGI}(a) \,\mathcal{O}_{\rm bare}\,
\label{RGI1}
\end{equation}
with
\begin{equation}
 \Delta Z^{\mathcal{S}}(M) = \left(2\beta_0 \frac{g^\mathcal{S}(M)^2}{16\,\pi^2}\right)^{-(\gamma_0/2\beta_0)}
 \, {\rm exp}\left\{ \int_0^{g^{\mathcal{S}}(M)} dg'   \left( \frac{\gamma^{\mathcal{S}}(g')}{\beta^{\mathcal{S}}(g')}+
 \frac{\gamma_0}{\beta_0 g'} \right)  \right\}
 \label{RGI2}
\end{equation}
and the RGI renormalization constant (depending on $a$ via the lattice coupling)
\begin{equation}
Z^{\rm RGI}(a) = \Delta Z^{\mathcal{S}}(M) \, Z^{\mathcal{S}}_{\rm bare}(M,a)\,.
\label{RGI3}
\end{equation}
Here $g^\mathcal{S}$, $\gamma^{\mathcal{S}}$ and $\beta^{\mathcal{S}}$ 
are the coupling constant, the anomalous dimension and the
$\beta$-function in scheme $\mathcal{S}$, respectively. Relations 
(\ref{RGI1}), (\ref{RGI2}) and (\ref{RGI3}) allow us to compute
the operator $\mathcal{O}$ in any scheme and at any scale we like,
once $Z^{\rm RGI}$ is known. Therefore, the knowledge of $Z^{\rm RGI}$ 
is very useful for the renormalization procedure in general.
Ideally, $Z^{\rm RGI}$ depends only on the bare lattice coupling,
but not on the momentum $p$.
Computed on a lattice, however, it suffers from lattice artifacts,
e.g., it contains contributions proportional to $a^2p^2$,
$(a^2p^2)^2$ etc. For a 
precise determination it is essential to have these discretization 
errors under control.

As the ${\rm RI}^\prime$-MOM scheme is in general not $O(4)$-covariant even in the
continuum limit, it is not very suitable for computing the anomalous
dimensions needed in (\ref{RGI2}).
Therefore we use an intermediate scheme $\mathcal{S}$ with known 
anomalous dimensions and calculate $Z^{\rm RGI}$ as follows:
\begin{equation}
Z^{\rm RGI}(a) = \Delta Z^{\mathcal{S}}(M=\mu)\, Z^{\mathcal{S}}_{\rm RI^\prime-MOM}(M=\mu) \, Z^{\rm RI'-MOM}_{\rm bare}(\mu,a)\,.
\label{RGI4}
\end{equation}
It turns out that a type of momentum subtraction scheme is a good
choice for $\mathcal S$ (for details see Ref.~\cite{Gockeler:2010yr}). The formula which 
is used to compute the transformation factor 
$Z^\mathcal{S}_{\rm RI'-MOM}(\mu)$ is given  
there together with all needed coefficients of 
the $\beta$-function and anomalous dimensions, which
are based on continuum three-loop calculations such as those in~\cite{Larin:1993vu,Retey:2000nq,Gracey:2006zr}.

On a lattice with linear extent $L$ the 
scale $\mu$ should ideally fulfill the relation 
\begin{equation}
1/L^2 \ll \Lambda^2_{\rm QCD} \ll \mu^2 \ll 1/a^2\,.
\label{eq:pts3}
\end{equation}
In that case $Z^{\rm {RGI}}(a)$ would be independent of $\mu$,
and from the resulting plateau we could read off the corresponding
final value. However, in practice $a \mu$ is not necessarily small
leading to non-negligible lattice artifacts that have to be tamed.
A promising tool to control lattice artifacts in a
systematic way is lattice perturbation theory: We expect that after
subtracting  these perturbative terms the calculation of the $Z$ factors can be
done more accurately.

\section{Subtraction of all lattice artifacts in one-loop order}
\label{sec:PTSUB1}

In standard lattice perturbation theory the one-loop renormalization
constants are given in the form
\begin{equation}
Z(\mu,a) = 1 + \frac{g^2\,C_F}{16\,\pi^2}\,\left(\gamma_0 \,\ln (a\mu) +  \Delta\right),\quad C_F=\frac{4}{3}\,.
\label{Zpert}
\end{equation}
This means that the $a$-dependence is retained only in the logarithm
and implicitly in $g$, while in all other contributions the limit
$a \to 0$ has been taken. 

However, there is no need to do so. 
We can keep $a$ finite everywhere and thus evaluate the lattice 
artifacts at one-loop order completely, proceeding as follows.
Let us denote by $F(p,a)$ the total one-loop correction and by
$\tilde{F}(p,a)$ the expression resulting from $F(p,a)$ by neglecting
all contributions which vanish for $a \to 0$. The difference
\begin{equation}
D(p,a) = F(p,a) - \tilde{F}(p,a)
\end{equation}
represents the lattice artifacts in one-loop perturbation theory
and is used to correct for the discretization errors:
\begin{equation}
Z_{\rm bare}^{\rm RI'-MOM}(p,a)_{\rm MC,sub} = Z_{\rm bare}^{\rm RI'-MOM}(p,a)_{\rm MC} - 
    \frac{g_{\star}^2}{16\,\pi^2} C_F \,D(p,a)\,.
 \label{Zsuball}
\end{equation}
There is a certain freedom in choosing the coupling $g_{\star}$ 
in (\ref{Zsuball}). It turned out that the use of the boosted coupling
\begin{equation}
 g_{\rm B}^2=\frac{g^2}{P(g)} = g^2 + O(g^4)
\label{gboost}
\end{equation}
($P(g)$ being the measured plaquette at $\beta=6/g^2$) 
is quite successful in estimating
the higher-order discretization effects.
With the prescription (\ref{Zsuball}) all lattice artifacts in 
one-loop order are subtracted.

\begin{figure}[!htb]
  \begin{center}
     \begin{tabular}{lcr}
        \includegraphics[scale=0.56,clip=true]
         {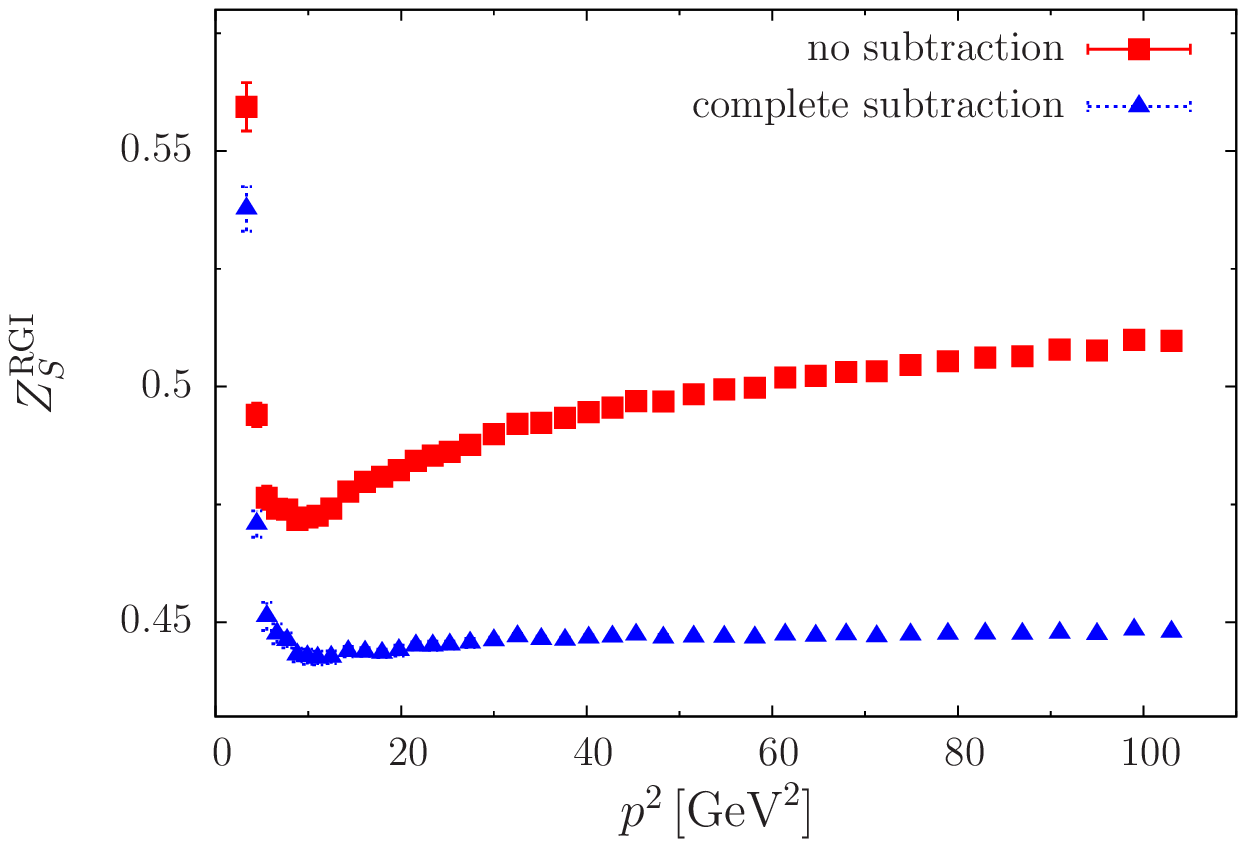}
&&
       \includegraphics[scale=0.56,clip=true]
         {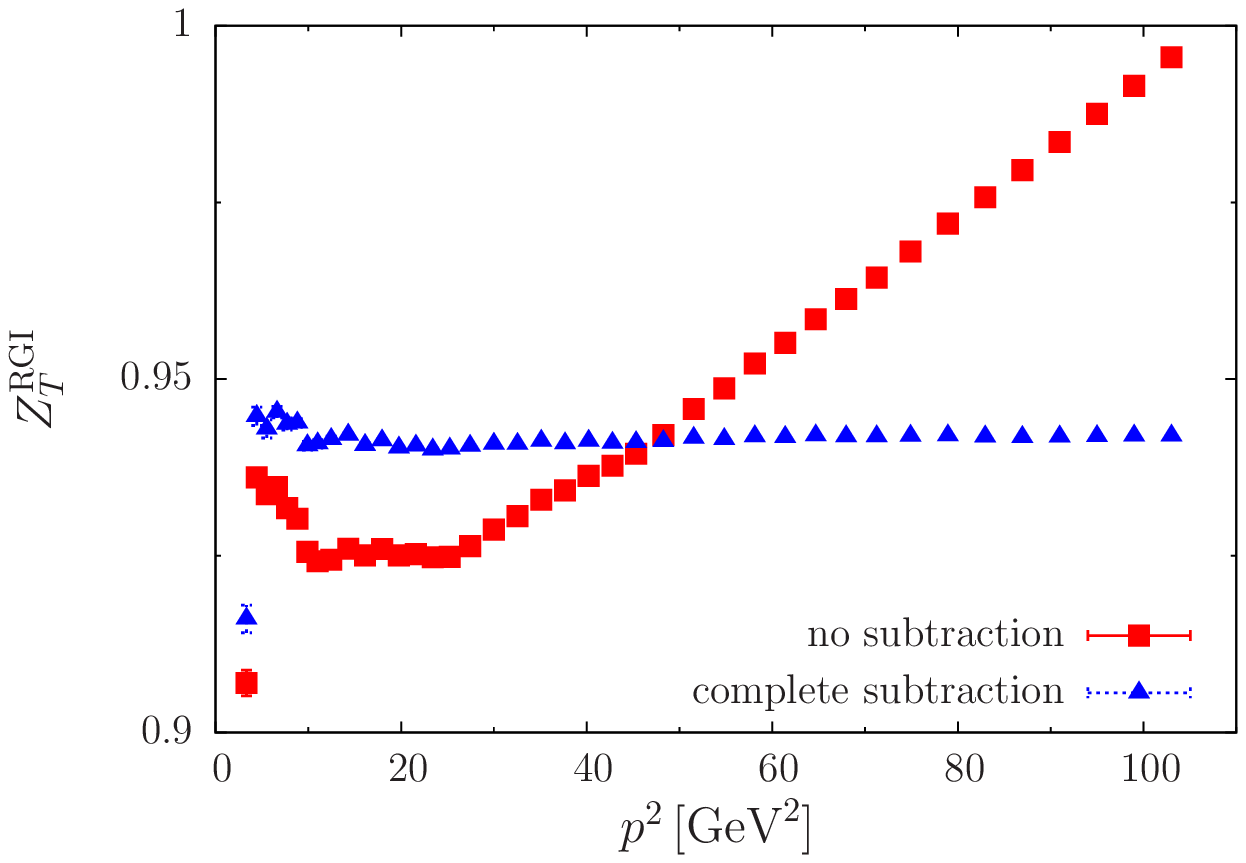}
     \end{tabular}
  \end{center}\vspace{-0.5cm}
  \caption{$Z_S^{\rm RGI}$  (left) and $Z_T^{\rm RGI}$ (right) for $\beta=5.40$. The $Z$ factors
obtained without subtraction are shown as red squares, those with complete one-loop subtraction (\ref{Zsuball})
as blue triangles. (The necessary scale transformation factors for the momenta are 
given at the end of Section~\ref{sec:PTSUB2}.)}
  \label{fig:pSTtotsub}
\end{figure}

In Fig.~\ref{fig:pSTtotsub} we show the effect of subtraction on the
RGI renormalization factors for selected operators of Table~\ref{OpTab}.
For all operators we recognize after subtraction a remarkable smoothing and a pronounced 
plateau as a function of $p^2$ for $p^2 \gtrsim  10\, {\rm GeV}^2$. The large 
bending in the small $p^2$ region might indicate the breakdown of 
perturbation theory (cf.\ the discussion in \cite{Gockeler:2010yr}).
The examples show that the one-loop subtraction of lattice artifacts (\ref{Zsuball}) works
very well and, moreover, is needed for a precise determination of 
the renormalization constants. The final values  for $Z^{\rm RGI}$ 
from (\ref{RGI4}) are obtained by a fit with an ansatz~\cite{Gockeler:2010yr}
\begin{equation}
Z^{\mathcal{S}}_{\rm RI'-MOM}(p) \, Z_{\rm bare}^{\rm RI'-MOM}(p,a)_{\rm MC,sub} = 
\frac{Z^{\rm RGI}(a)}{ \Delta Z^{\mathcal{S}}(p)\,\left[1+b_1\,(g^\mathcal{S})^8\right]}+c_1\, a^2 p^2\,.
\label{RGI4n}
\end{equation}
The free parameter $b_1$  takes into account that the transformation 
factor $Z^{\mathcal{S}}_{\rm RI'-MOM}(p)$ is known to three-loop 
order $\left(g^\mathcal{S}\right)^6$ only. Further possible 
lattice artifacts are parametrized by $c_1\, a^2 p^2$.

For practical reasons the numerical calculation of $F(p,a)$ - and therefore the
calculation of $Z^{\rm RGI}$ using (\ref{RGI4n}) - is restricted to  
operators with at most one derivative and for $N_f = 2$ only. In order to perform the 
subtraction for a wider class of operators and/or for $N_f = 2+1$ 
(where the considered lattice action becomes more complicated) 
we have to look for an  alternative method.
One possibility which will be discussed in the next sections is a
'reduced' subtraction: Instead of subtracting the complete
one-loop lattice artifacts we subtract only the one-loop 
terms proportional to $a^2$, if they are known for the given action.

\section{Subtraction of order $a^2$ one-loop lattice artifacts}
\label{sec:PTSUB2}

\subsection{Lattice perturbation theory up to order  $g^2a^2$}
\label{sec:PTSUB21}

The diagrammatic approach to compute the one-loop $a^2$ terms for the 
$Z$ factors of local and one-link operators has been developed
by some of us~\cite{Constantinou:2009tr,Skouroupathis:2010zz}. 
The general case of Wilson type improved fermions is discussed 
in~\cite{Alexandrou:2012mt}. For details of the computations we 
refer to these references. Here we give explicitly the results for 
the operators and actions investigated in this paper (massless 
improved Wilson fermions with $\csw=1$, plaquette gauge action,
Landau gauge). 

Using the relation (\ref{ZRinv})  we compute a common $Z$ factor for each
multiplet given in Table~\ref{OpTab}. 
The results are as follows:
\vspace{0.8cm}
\begin{eqnarray}
Z_S &=& 1 + \frac{g^2\,C_F}{16 \pi^2}\,
\Bigg\{-23.3099 + 3\,\log(a^2S_2) \nonumber\\
&&+ a^2\left[S_2\left(1.64089 - \frac{239}{240} \log (a^2{S_2})\right)+
     \frac{S_4}{S_2}\left(1.95104 - \frac{101}{120}\log (a^2{S_2})\right)\right]\Bigg\}\,,
\nonumber\\
Z_V &=& 1 + \frac{g^2\,C_F}{16 \pi^2}\,
\Bigg\{-15.3291  \nonumber\\
&& + a^2\left[S_2\left(-1.33855+ \frac{151}{240} \log (a^2{S_2})\right)+
     \frac{S_4}{S_2}\left(2.89896 - \frac{101}{120}\log (a^2{S_2})\right)\right]\Bigg\}\,,
\nonumber\\
Z_A &=& 1 + \frac{g^2\,C_F}{16 \pi^2}\,
\Bigg\{-13.7927  \nonumber\\
&&+ a^2\left[S_2\left(-0.92273 + \frac{151}{240} \log (a^2{S_2})\right)+
     \frac{S_4}{S_2}\left(2.89896 - \frac{101}{120}\log (a^2{S_2})\right)\right]\Bigg\}\,,
\nonumber\\
Z_T &=& 1 + \frac{g^2\,C_F}{16 \pi^2}\,
\Bigg\{-11.1325 - \,\log(a^2S_2) \label{Za2all}\\
&& +a^2\Bigg[S_2\left(-1.72760 + \frac{221}{240} \log (a^2{S_2})\right)+
     \frac{S_4}{S_2}\left(3.21493 - \frac{101}{120}\log (a^2{S_2})\right) \Bigg]\Bigg\}\,,
\nonumber\\
Z_{v_{2,a}} &=&   1 + \frac{g^2\,C_F}{16 \pi^2}\,
\Bigg\{ 6.93831 -
      \frac{8}{3}\log (a^2 S_2) -
      \frac{2}{9}\frac{S_4}{(S_2)^2} \nonumber \\
&&+{a^2}\,\Bigg[
          S_2\,\left( -1.50680 +
            \frac{167}{180} \log (a^2S_2)\right)
\nonumber \\
&&\hspace{0.3cm} +\frac{S_4}{S_2}\,\left( 2.63125 -
              \frac{197}{180}\log (a^2S_2) \right)-\frac{71}{540}\frac{{S_4}^2}{(S_2)^3} -
         \frac{82}{135}\frac{S_6}{(S_2)^2} \Bigg]\Bigg\}\,,\nonumber
\nonumber\\
Z_{v_{2,b}} &=& 1 +  \frac{g^2\,C_F}{16 \pi^2}\,
\Bigg\{  5.78101  -
      \frac{8}{3}\log (a^2S_2)+
      \frac{4}{9}\frac{S_4}{(S_2)^2}\nonumber \\
&&+{a^2}\,\Bigg[ 
         S_2\,\left( -0.56888 +
            \frac{1}{30}\log (a^2S_2) \right)
\nonumber \\
&&\hspace{0.3cm} +\frac{S_4}{S_2}\,\left( -0.51323 +
              \frac{19}{30}\log (a^2S_2) \right)+\frac{71}{270}\frac{{S_4}^2}{(S_2)^3} +
         \frac{164}{135}\frac{S_6}{(S_2)^2} \Bigg]\Bigg\}\,.\nonumber 
\nonumber
\end{eqnarray}
Here we have introduced the notation
\begin{equation}
  S_n = \sum_{\lambda=1}^4 \, p_\lambda^n\,, 
  \label{Sn}
\end{equation}
with $p_\lambda$ being the momentum components. 
Note that terms of type $(S_4/S_2)\log(a^2 S_2)$, appearing in $Z_S,\
Z_V,\ Z_A,\ Z_T$, all have the same coefficient which
arises solely from the quark wave function renormalization constant $Z_q$.
The corresponding one-loop vertex functions $\Gamma_i(p)$ in (\ref{ZRinv})
do not contain such a structure.
For later purposes we write the $Z$ factors generically as
\begin{equation}
 Z = 1 + \frac{g^2\,C_F}{16 \pi^2}\,Z_{\rm 1-loop} + {a^2} g^2 Z^{(a^2)}_{\rm 1-loop}(p,a)\,.
\label{Za2corr}
\end{equation}
We emphasize that the numerical coefficients in the above expressions 
are either exact rationals or can be computed to a very high precision.

\begin{figure}[!htb]
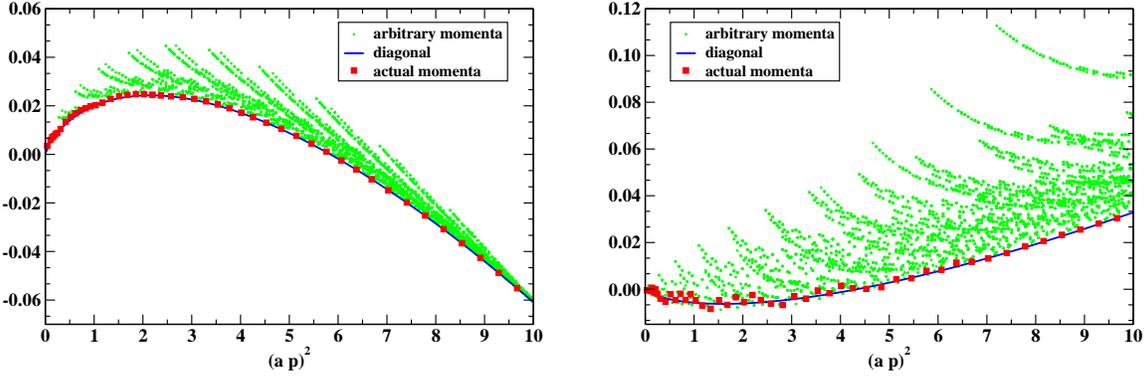

  \begin{center}
     \begin{tabular}{lcr}
        \includegraphics[scale=0.3,clip=true]
         {Figures/Zs_alter.eps}
&&
       \includegraphics[scale=0.3,clip=true]
         {Figures/Zv_alter.eps}
     \end{tabular}
  \end{center}\vspace{-0.5cm}
  \caption[]{$a^2 g^2 Z^{(a^2)}_{\rm 1-loop}(p,a)$ for operators 
$\mathcal{O}^{S}$ (left) and $\mathcal{O}^{V}$ (right) as a 
function of $a^2p^2$ on a $24^3 \times 48$ lattice at $\beta=5.40$. The green filled circles are 
the values for an arbitrary set of (mostly non-diagonal) momenta,
whereas the red filled squares are obtained from the momenta used in this 
investigation. The blue line is computed from diagonal momenta.}
  \label{fig:ZSVa2corr}
\end{figure}

\begin{figure}[!htb]
\vspace{0.5cm}
  \begin{center}
     \begin{tabular}{lcr}
        \includegraphics[scale=0.3,clip=true]
         {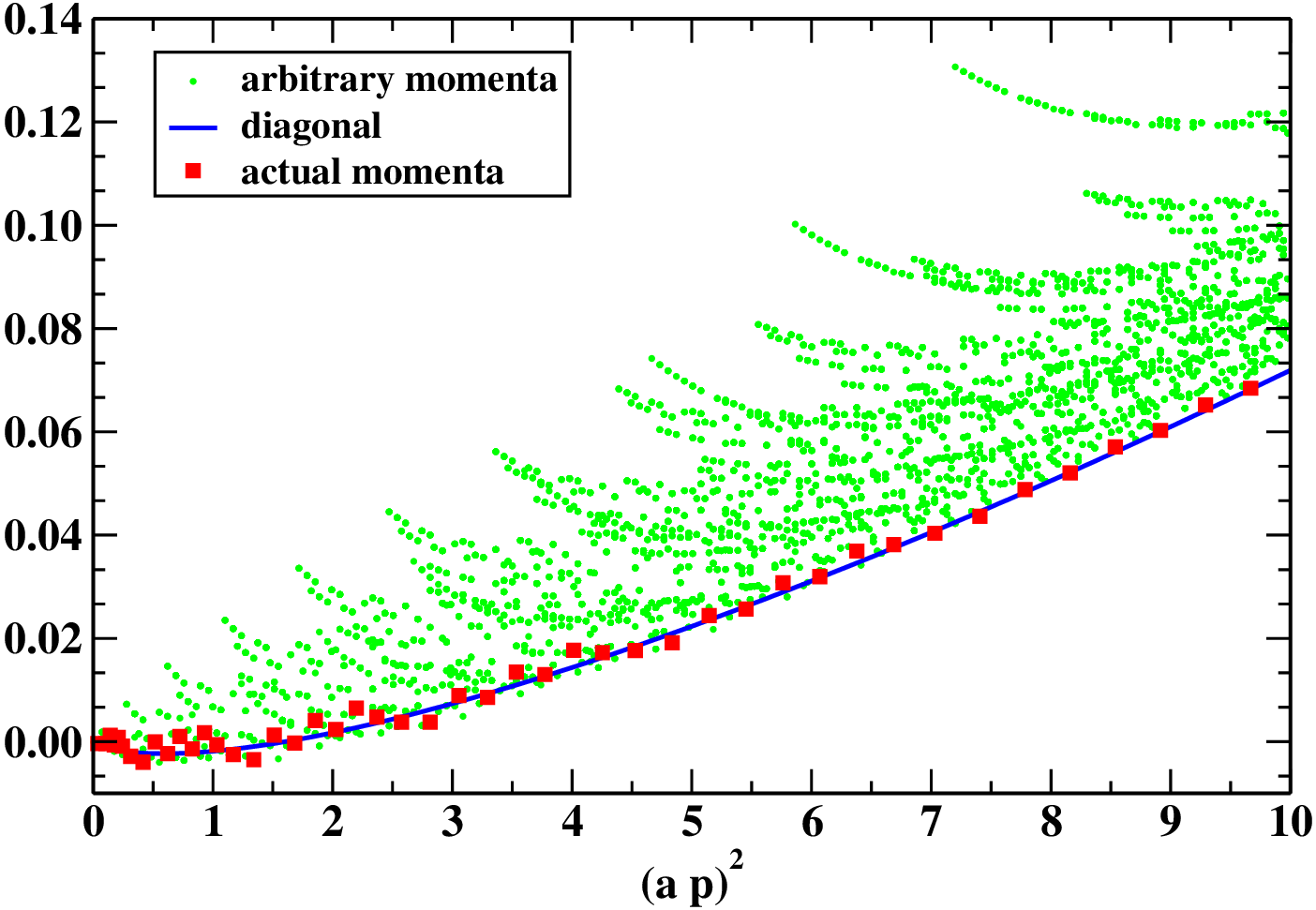}
&&
       \includegraphics[scale=0.3,clip=true]
         {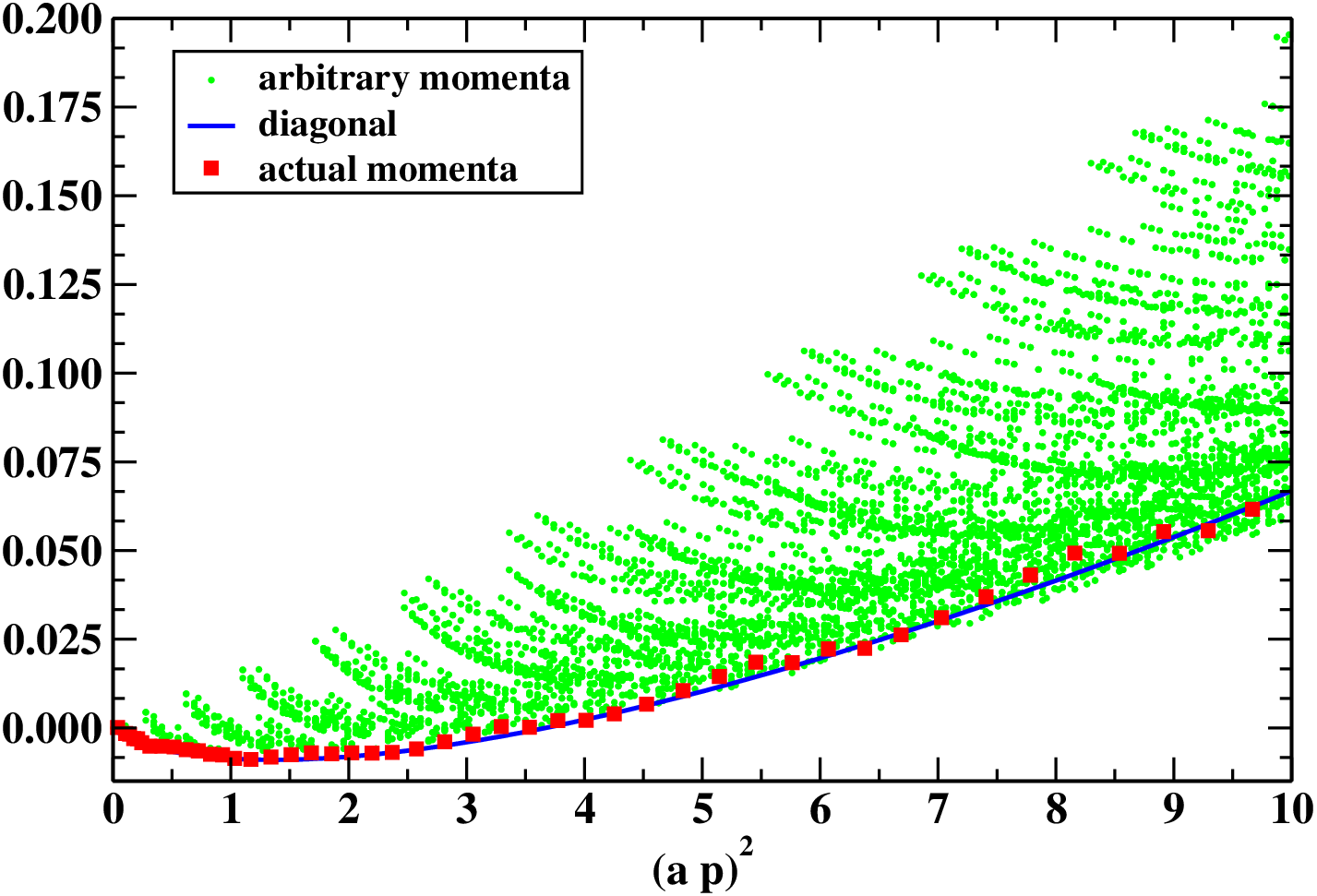}
     \end{tabular}
  \end{center}\vspace{-0.5cm}
  \caption[]{The same as Fig.~\ref{fig:ZSVa2corr} but for operators $\mathcal{O}^{A}$ (left) and
$\mathcal{O}^{T}$ (right).}
  \label{fig:ZATa2corr}
\end{figure}

\begin{figure}[!htb]
\vspace{0.5cm}
  \begin{center}
     \begin{tabular}{lcr}
        \includegraphics[scale=0.3,clip=true]
         {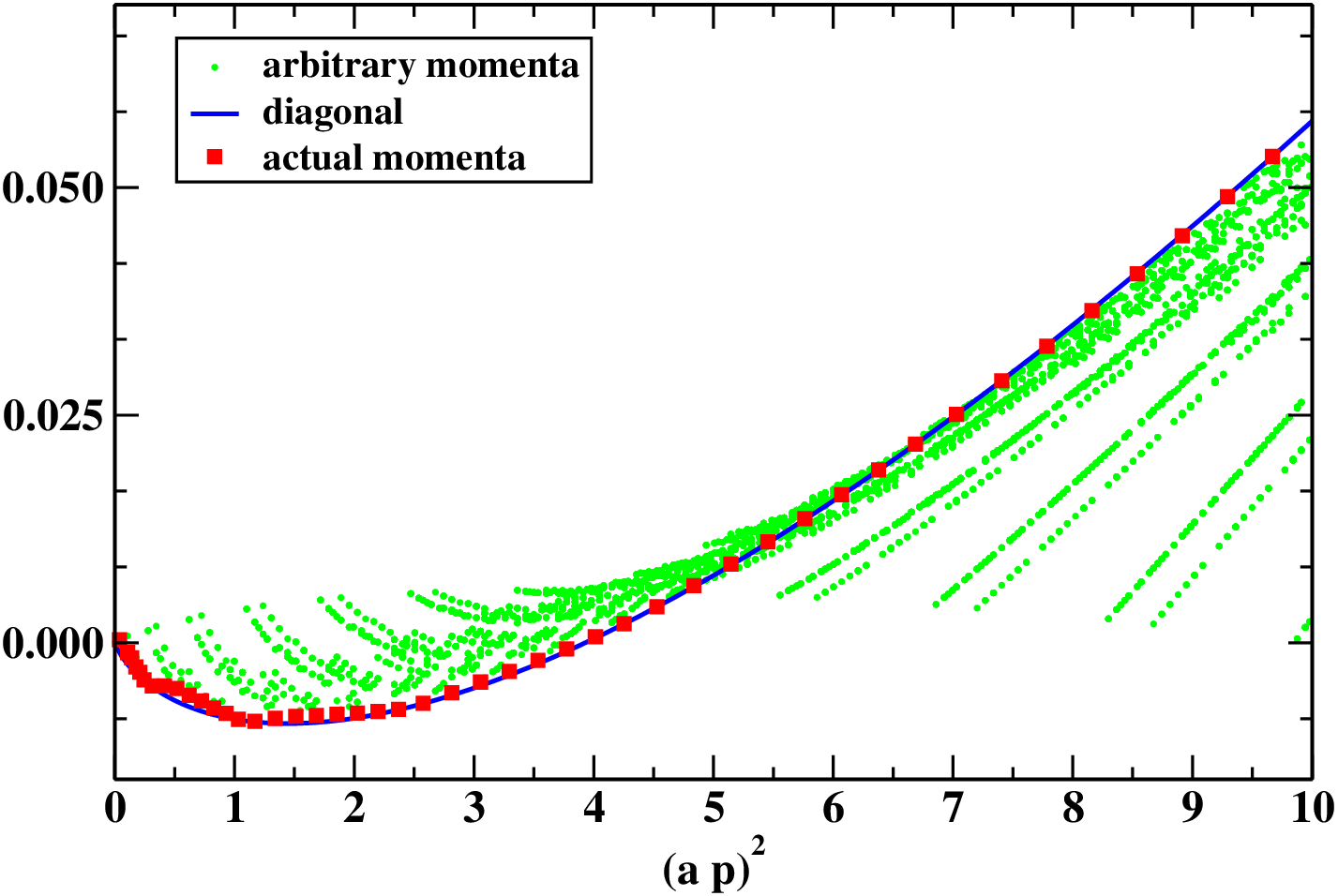}
&&
       \includegraphics[scale=0.3,clip=true]
         {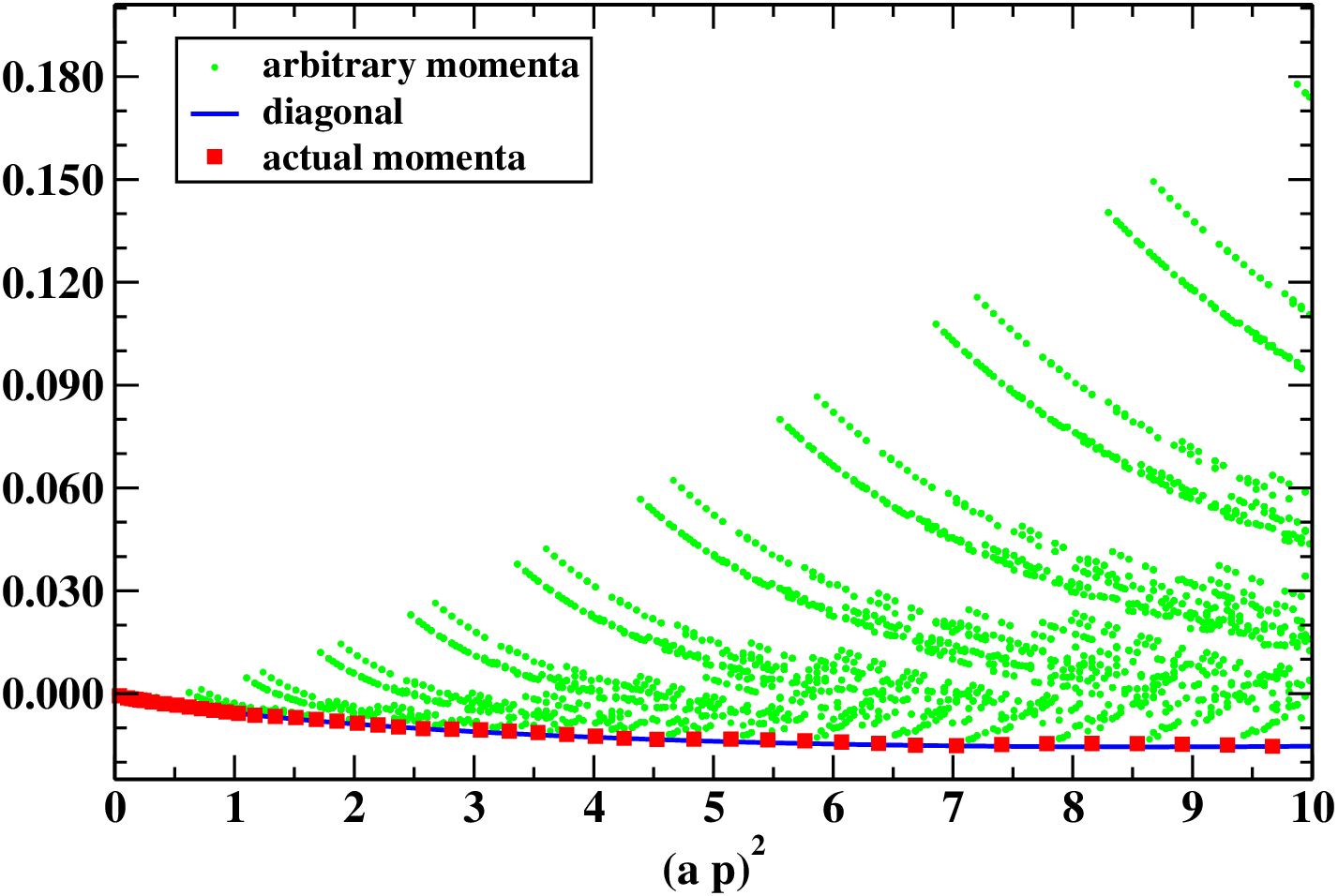}
     \end{tabular}
  \end{center}\vspace{-0.5cm}
  \caption[]{The same as Fig.~\ref{fig:ZSVa2corr} but for operators $\mathcal{O}^{v_{2,a}}$ (left) and
$\mathcal{O}^{v_{2,b}}$ (right).}
  \label{fig:Zv2av2ba2corr}
\end{figure}

In Figs.~\ref{fig:ZSVa2corr}, \ref{fig:ZATa2corr} and \ref{fig:Zv2av2ba2corr} we present 
$ {a^2} g^2 Z^{(a^2)}_{\rm 1-loop}(p,a)$
for selected operators as a function of $a^2p^2$
on a finite lattice, where we choose the lattice momenta as
$p_\lambda=({2\pi\,i_\lambda})/({a\,L_\lambda})$.
Here, $i_\lambda$ are integers and  $L_\lambda$ is the lattice extension in direction $\lambda$.
We compare the correction terms for a general set of momenta with
those obtained for the momenta used in this investigation at 
$\beta = 5.40$ on $24^3 \times 48$ lattices and with 
'diagonal' momenta, i.e., momenta on the diagonal of the Brillouin 
zone.

\newpage\clearpage
The figures show that the momenta of the actually measured $Z$ factors 
are very close to the diagonal. Furthermore, one recognizes that the 
magnitude of the calculated  one-loop $a^2$ corrections 
in the used  momentum range is small but not negligible compared
to the measured values which are of order $1$ 
(see also Fig.~\ref{fig:pSTtotsub}).
Therefore, one can expect that the subtraction of those terms yields a noticeable effect.

\subsection{Subtraction of lattice artifacts up to order  $a^2$}
\label{sec:PTSUB22}
The subtraction procedure of order $a^2$ terms is not unique - we can use different definitions. 
The only restriction is
that at one-loop order they should agree
(treating $Z^{\rm RI'-MOM}_{\rm bare}(p,a)_{\rm MC}$ in perturbation theory).
We investigate the following possibilities,
\begin{eqnarray}
Z^{\rm RI'-MOM}_{\rm bare}(p,a)_{\rm MC,sub,s} &=& Z^{\rm RI'-MOM}_{\rm bare}(p,a)_{\rm MC} - 
 {a^2} \, g_\star^2 \, Z^{(a^2)}_{\rm 1-loop}(p,a)\,, \label{eq:pts11} \\
Z^{\rm RI'-MOM}_{\rm bare}(p,a)_{\rm MC,sub,m} &=& Z^{\rm RI'-MOM}_{\rm bare}(p,a)_{\rm MC}\, \times
  \left(1 -{a^2} \, g_\star^2 \, Z^{(a^2)}_{\rm 1-loop}(p,a)\right)\,, \label{eq:pts12} 
\end{eqnarray}
where $g_\star$ can be chosen to be either the bare lattice coupling $g$ 
or the boosted coupling $g^{}_{\rm B}$ (\ref{gboost}).
(In the following we denote subtraction type (\ref{eq:pts11}) by ({\bf s}) and
(\ref{eq:pts12}) by ({\bf m})).
With ansatz ({\bf s}) the one-loop $a^2$ correction is subtracted 'directly'
from $Z^{\rm RI'-MOM}_{\rm bare}(p,a)_{\rm MC}$.
Subtraction type ({\bf m}) factorizes the one-loop $a^2$ correction from the 
nonperturbative $Z$ factor.

The $Z^{\rm RGI}$ are computed from (\ref{RGI4}) using ({\bf s}) or ({\bf m}),
where we expect slightly different numbers depending on the choice of coupling $g_\star$.
The only significant errors to $Z^{\rm RI'-MOM}_{\rm bare}(p,a)_{\rm MC,sub}$ 
are due to the Monte Carlo simulations.

\begin{figure}[!htb]
  \begin{center}
     \begin{tabular}{lcr}
        \includegraphics[scale=0.56,clip=true]
         {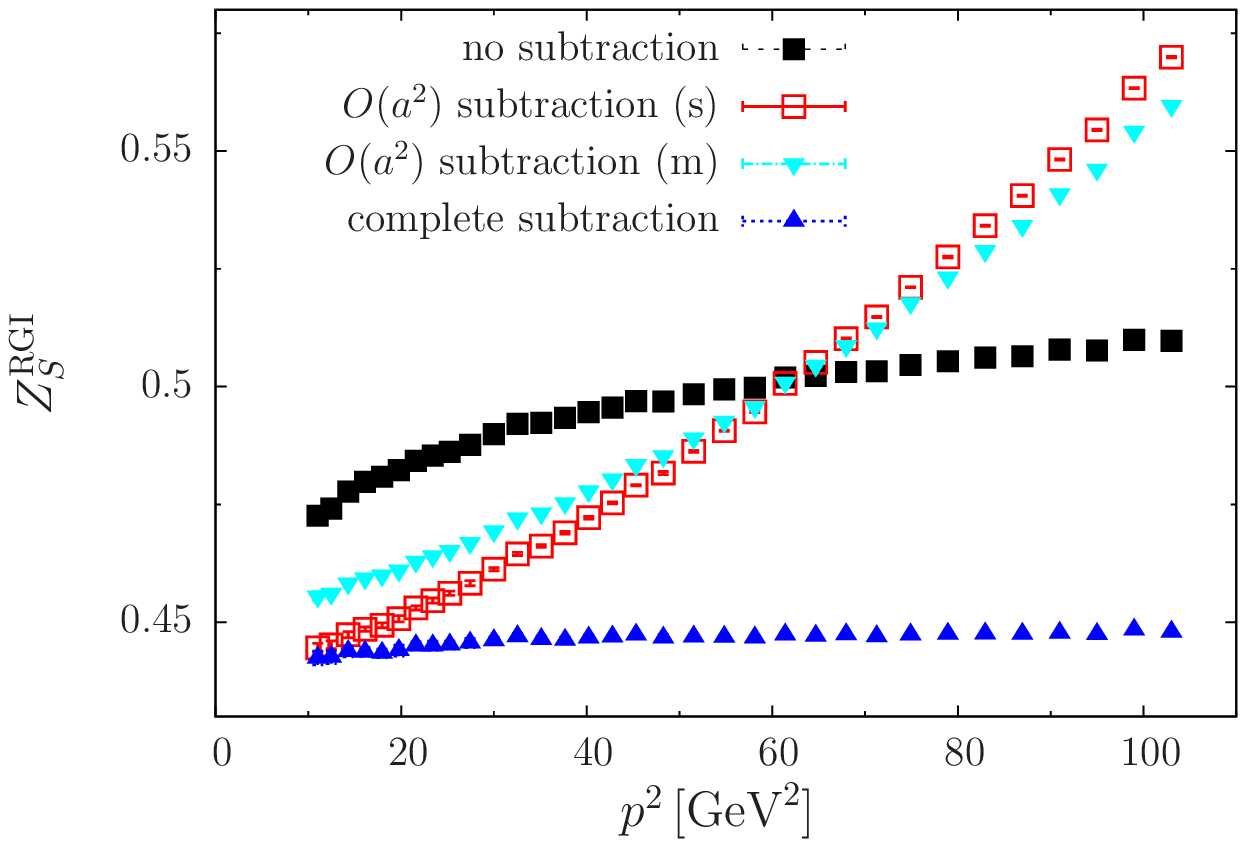}
&&
       \includegraphics[scale=0.56,clip=true]
         {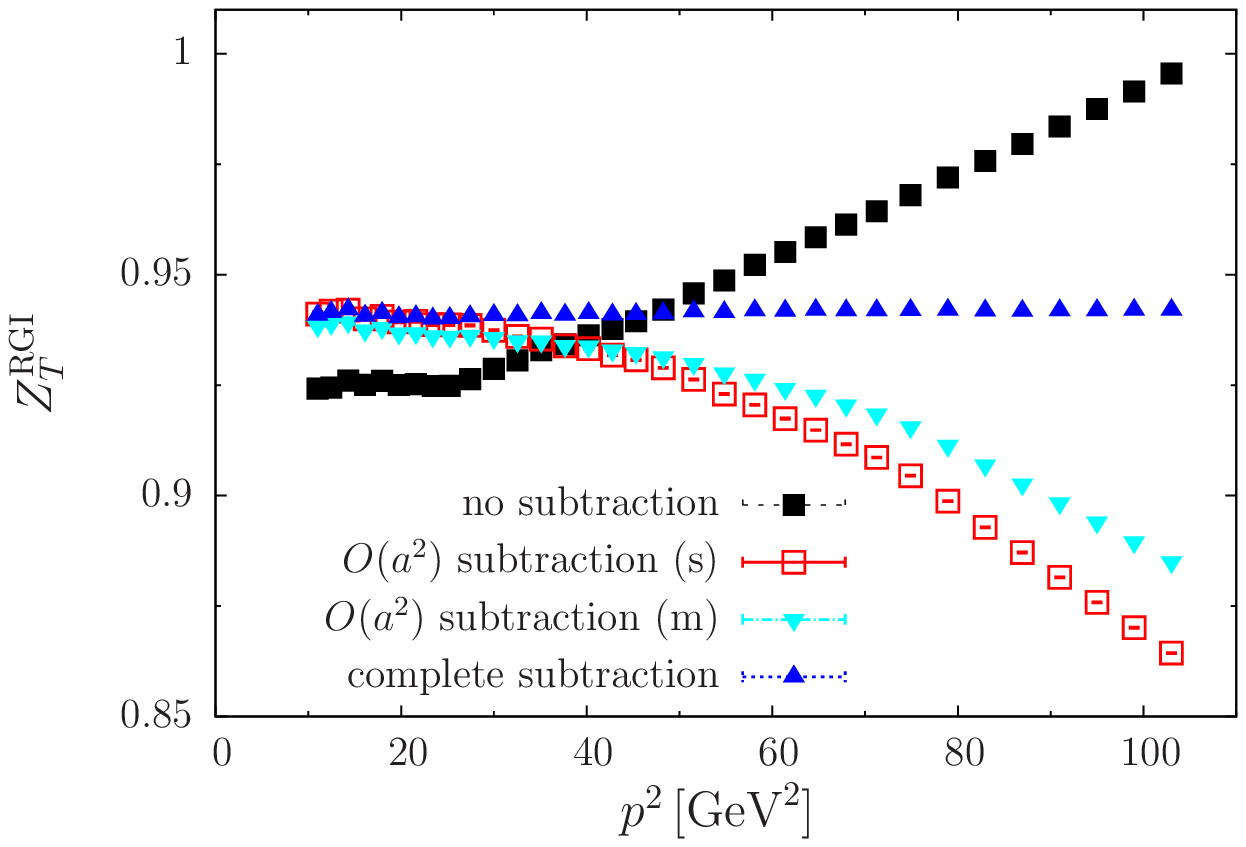}
     \end{tabular}
  \end{center}\vspace{-0.5cm}
  \caption[]{Unsubtracted and subtracted renormalization constants for the scalar operator $\mathcal{O}^S$ (left)
and the tensor operator $\mathcal{O}^T$ (right) at $\beta=5.40$, for $p^2 \gtrsim 10\, {\rm GeV}^2$
and $r_0\,\Lambda_{\rm \overline{MS}}=0.700$. The complete subtraction is based on (\ref{Zsuball}), whereas
the $a^2$ subtractions are of type ({\bf s}) and ({\bf m}) with $g_\star=g^{}_{\rm B}$.}
  \label{fig:ZSZT}
\end{figure}

In Fig.~\ref{fig:ZSZT} we show 
how  the subtraction of lattice artifacts (complete and $a^2$) 
affects the renormalization constants for the scalar and tensor operators.
The complete one-loop subtraction  results in a clear plateau for both
$Z^{\rm RGI}$ factors. Using the $a^2$ subtractions there remains a more or less
pronounced curvature which has to be fitted. 
From the definitions of the subtraction terms it is clear that they
vanish at $a^2 p^2 = 0$. Moreover, for small $p^2 \approx 10 \, {\rm GeV}^2$ the subtraction
methods ({\bf s}) and (\ref{Zsuball}) already agree, as they should. 
However, as discussed above, $Z^{\rm {RGI}}$ can only be determined 
from sufficiently large momenta ($p^2 \gtrsim 10 \, \mbox{GeV}^2$), 
where  differences arise between the various procedures. 
Therefore the results for $Z^{\rm {RGI}}$ may differ depending on the kind of
subtraction. As can be seen in Fig.~\ref{fig:ZSZT}, this effect varies 
strongly from operator to operator.

\subsection{Fit procedure}

Compared to the complete one-loop subtraction we expect that $Z^{\rm RI'-MOM}_{\rm bare}(p,a)_{\rm MC,sub}$ as
computed from ({\bf s}) or ({\bf m}) 
contains terms proportional to 
$a^{2n}$ ($n \geq 2$) even at order $g^2$, as well as the lattice
artifacts from higher orders in perturbation theory, constrained
only by hypercubic symmetry. Therefore, we parametrize the subtracted 
data for each $\beta$ in terms of the hypercubic invariants $S_n$
defined in (\ref{Sn}) as follows
\begin{eqnarray}
&& Z^\mathcal{S}_{\rm RI'-MOM}(p) \,
			Z^{\rm RI'-MOM}_{\rm bare}(p,a)_{\rm MC,sub} = 
  \frac{Z^{\rm RGI}(a)}{ \Delta Z^{\mathcal{S}}(p)\,\left[1+b_1\,(g^\mathcal{S})^8\right]} +
\label{struc1}
\\ 
&&
\hspace{-2mm}
a^2 \left( c_1\,S_2 + c_2 \,\frac{S_4}{S_2} +  c_3 \,\frac{S_6}{(S_2)^2}\right)
 + a^4 \, \left( c_4\, (S_2)^2 + c_5\, S_4 \right)
 +a^6 \left(c_6\, (S_2)^3 + c_7\, S_4 \,S_2 + c_8 \,S_6\right) \,.\nonumber
 \nonumber
\end{eqnarray}
There are also further non-polynomial invariants at order
$a^4, a^6$, but their behavior is expected to be well
described by the invariants which have been included already.
Ansatz (\ref{struc1}) is a generalization of (\ref{RGI4n}): 
After the 'reduced' one-loop subtraction of lattice artifacts the $Z$ factors are expected 
to depend more strongly on $a^4$ or $a^6$ hypercubic invariants 
than after the complete one-loop subtraction (see Fig.~\ref{fig:ZSZT}).
The parameters $c_1, \dots , c_8$ describe the lattice artifacts. 

Together
with the target parameter $Z^{\rm RGI}(a)$ we have ten parameters  for this general case.
In view of the limited number of data points for each single $\beta$ value
$(5.20$, $5.25$, $5.29$, $5.40)$ we apply the ansatz (\ref{struc1})  
to several $\beta$ values simultaneously with
\begin{equation}
\frac{Z^{\rm RGI}(a)}{ \Delta Z^{\mathcal{S}}(p)\,\left[1+b_1\,(g^\mathcal{S})^8\right]}
\to \frac{Z^{\rm RGI}(a_k)}{ \Delta Z_k^{\mathcal{S}}(p)\,\left[1+b_1\,(g^\mathcal{S})^8\right]}
  \,,\label{struc2}
\end{equation}
where $k$ labels the corresponding $\beta$ value ($a_k=a(\beta_k)$).
The parameters $c_i$ are taken to be independent of $\beta$.
This enhances the ratio (number of data points)/(number of fit parameters) significantly
and we obtain several $Z^{\rm {RGI}} (a_k)$ at once. The fit is performed by a
nonlinear model fit which uses - depending on the actual convergence - either the Nelder-Mead
or a differential evolution algorithm~\cite{mathematica}. Additionally, we have checked some of the fit results
using MINUIT~\cite{minuit}.

The renormalization factors are influenced by the choice for $r_0\,\Lambda_{\rm \overline{MS}}$.
This quantity enters $\Delta Z^{\mathcal{S}}(M)$ in  (\ref{RGI2}) via the corresponding coupling
$g^{\mathcal{S}}(M)$ (for details see~\cite{Gockeler:2010yr}). 
We choose $r_0\,\Lambda_{\rm \overline{MS}}=0.700$~\cite{QCDSF:2013}.
In order to estimate the influence of the choice of $r_0\,\Lambda_{\rm \overline{MS}}$
we also use $r_0\,\Lambda_{\rm \overline{MS}}=0.789$ calculated in~\cite{Fritzsch:2012wq}.
The Sommer scale $r_0$ is chosen to be $r_0=0.501 \, {\rm fm}$ and the relation between the lattice spacing $a$
and the inverse lattice coupling $\beta$ is given by $r_0/a =6.050\,(\beta=5.20), 6.603\,(\beta=5.25),
7.004\,(\beta=5.29)$ and $8.285\,(\beta=5.40)$~\cite{Bali:2012qs}.

\section{Renormalization factors for local and one-link operators}

The fit procedure as sketched above has quite a few degrees of freedom
and it is essential to investigate their influence carefully.
A criterion for the choice of the minimal value of $p^2$ is provided
by the breakdown of perturbation theory at small momenta. The data
suggest~\cite{Gockeler:2010yr} that we are on the 'safe side' when choosing
$p^2_{\min} = 10 \, \mbox{GeV}^2$. As the upper end of the fit
interval we take the maximal available momentum at given coupling
$\beta$. 

Other important factors are

\begin{itemize}

\item {\bf Type of subtraction: } As discussed above the procedure of the
one-loop subtraction is not unique. We choose different definitions
({\bf s}) and ({\bf m}) with either bare $g$ or
boosted coupling $g_B$. 

\item {\bf Selection of hypercubic invariants:}
For the quality of the fit it is essential how well we describe the 
lattice artifacts which remain after subtraction~\cite{Boucaud:2003dx,deSoto:2007ht}. 
This is connected to the question whether the $a^2$ subtraction has 
been sufficient to subtract (almost) all $a^2$ artifacts. Therefore, we perform
fits with various combinations of structures with coefficients $c_i$ in (\ref{struc1}).
One should mention that the concrete optimal (i.e.\ minimal) set of  $c_i$ depends
strongly on the momenta of the available Monte Carlo data - nearly diagonal
momenta require fewer structures to be fitted than far off-diagonal ones.

\end{itemize}

The analysis should provide an optimal restricted set of parameters which can be used as
a guideline for other classes of operators. Nevertheless, one has to inspect every
new case carefully.

The results for $Z^{\rm {RGI}}$ will depend on the above mentioned
factors. As a detailed presentation for all operators and $\beta$-values would be too
lengthy, we select some operators and/or $\beta$ values and take the corresponding results
as a kind of reference. All results presented
in this section are computed for 
$r_0\,\Lambda_{\rm \overline{MS}}=0.700$. The choice 
$r_0\,\Lambda_{\rm \overline{MS}}=0.789$ leads to qualitatively
similar results. The large number of parameters in ansatz (\ref{struc1})
calls for a combined use of the data sets at
$\beta=(5.20,5.25,5.29,5.40)$
for our fit analysis as indicated in (\ref{struc2}). With the choice $p^2_{\min} = 10 \, \mbox{GeV}^2$
this results in $94$  data points available for the corresponding fits. 
Additionally, we should note that the errors on our fit parameters
are those obtained from the nonlinear model fit. They differ
from the error calculation for the $Z^{\rm RGI}$ based on
 (\ref{RGI4n}) and used in~\cite{Gockeler:2010yr}.

\subsection{Dependence on the subtraction type}

\begin{figure}[!thb]
  \begin{center}
     \begin{tabular}{lcr}
        \includegraphics[scale=0.56,clip=true]
         {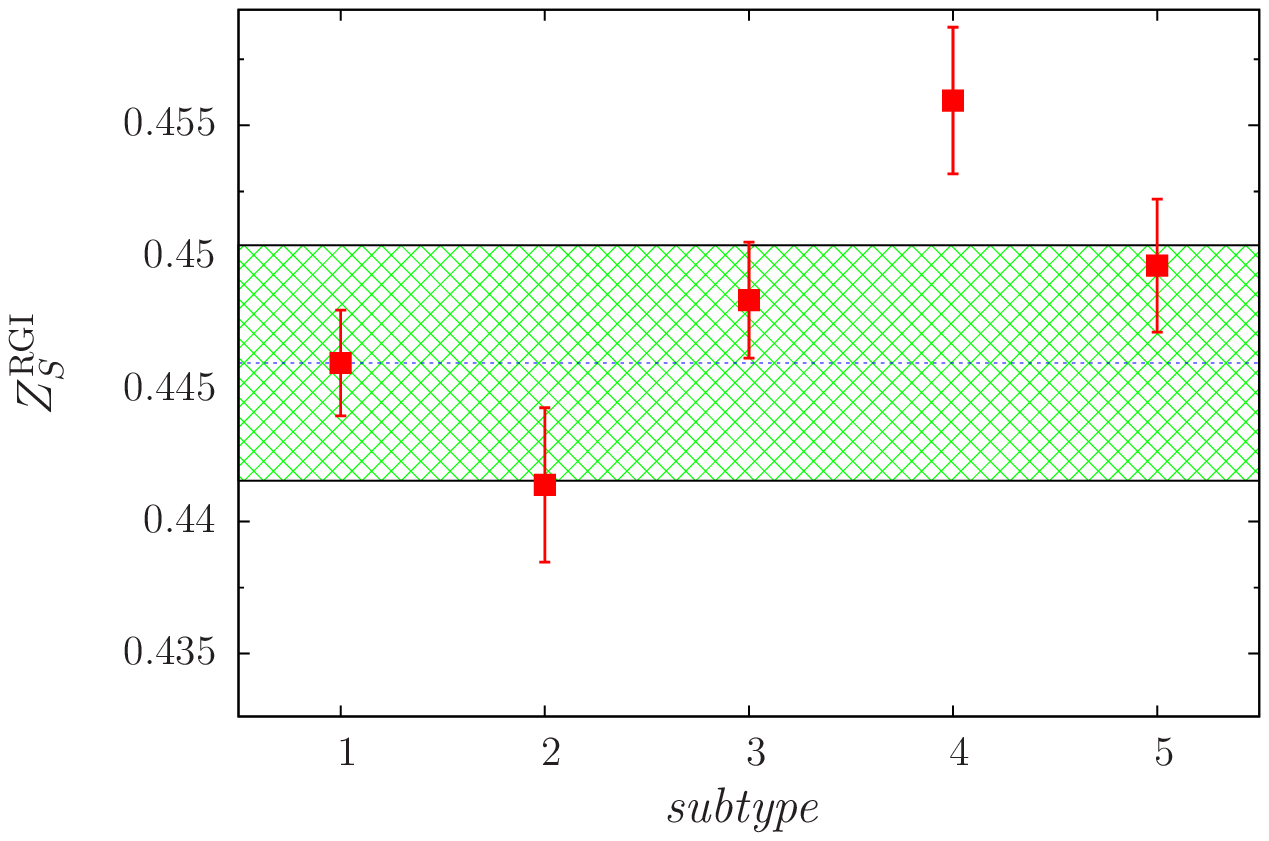}
&&
       \includegraphics[scale=0.56,clip=true]
         {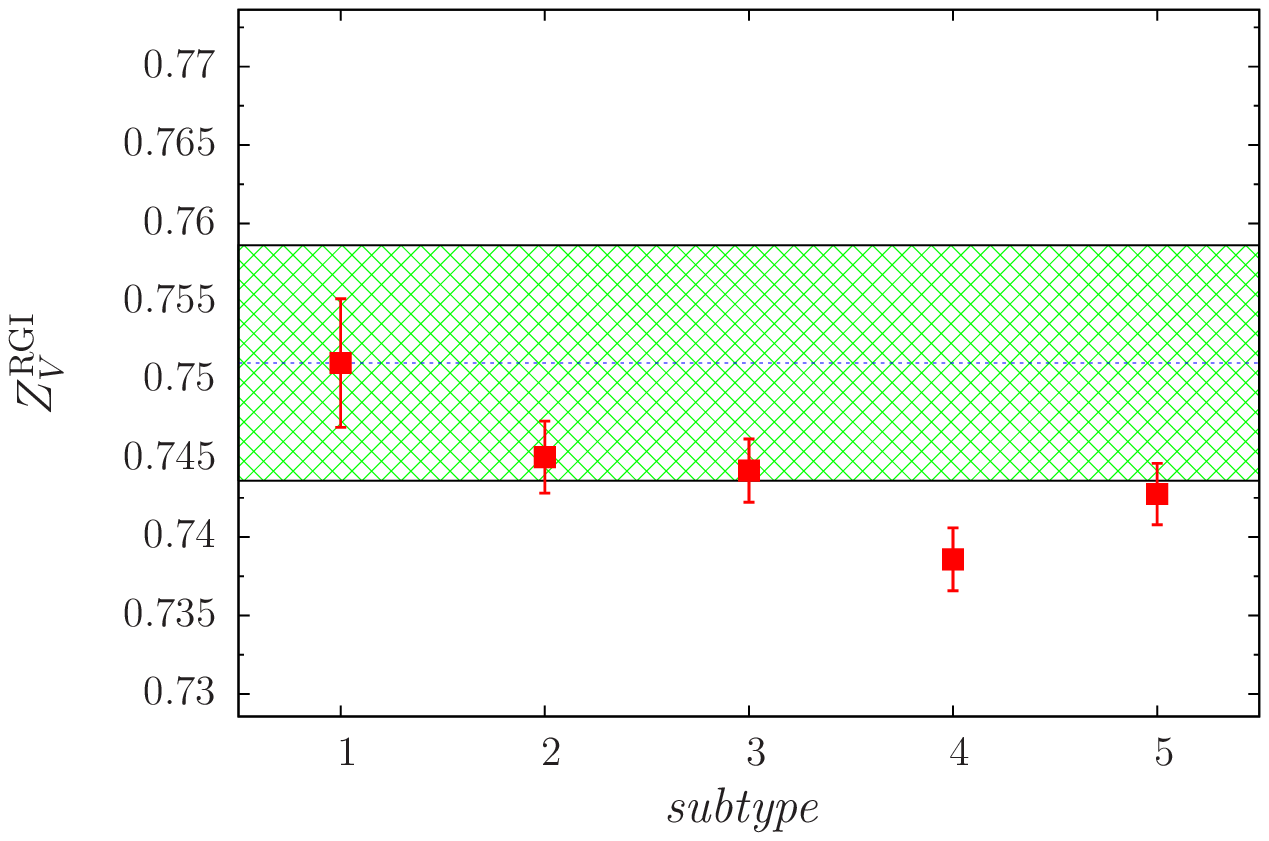}
\\
        \includegraphics[scale=0.56,clip=true]
         {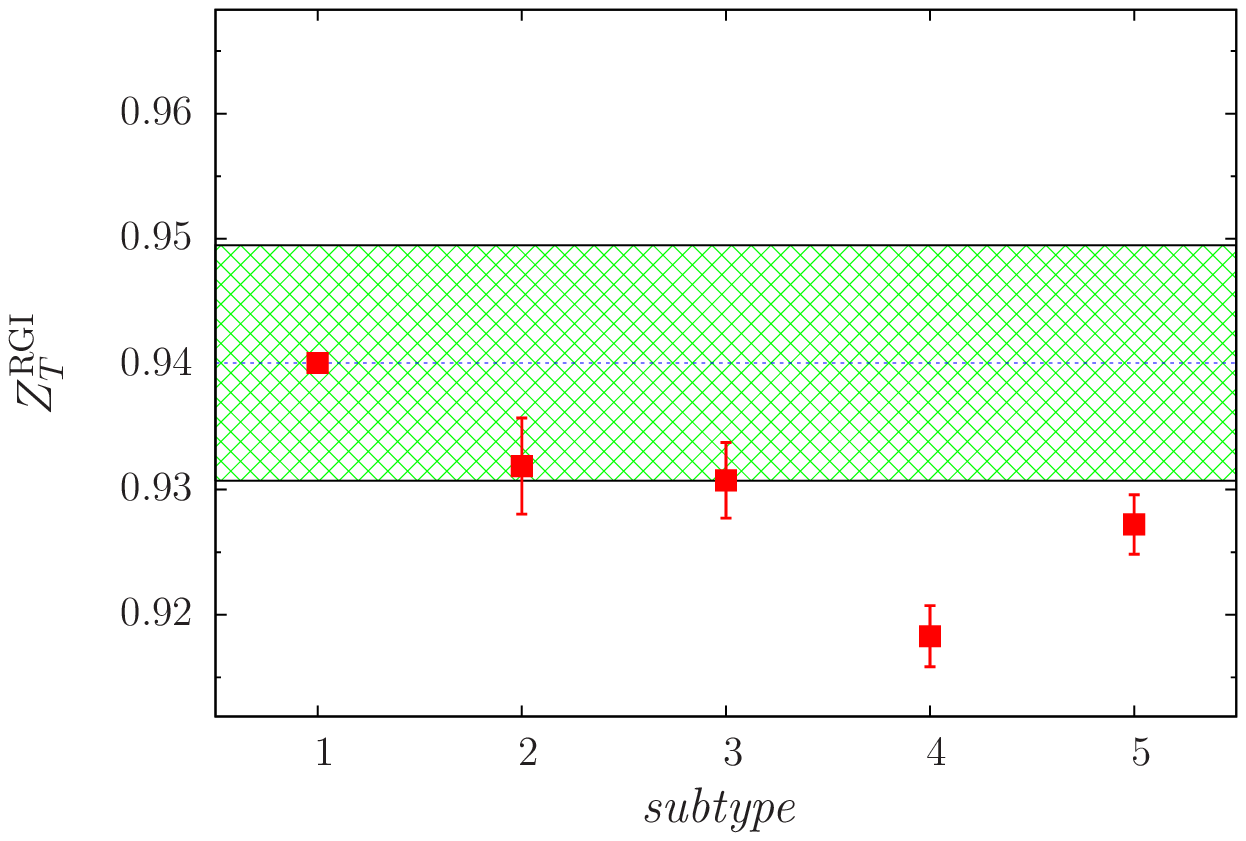}
&&
       \includegraphics[scale=0.56,clip=true]
         {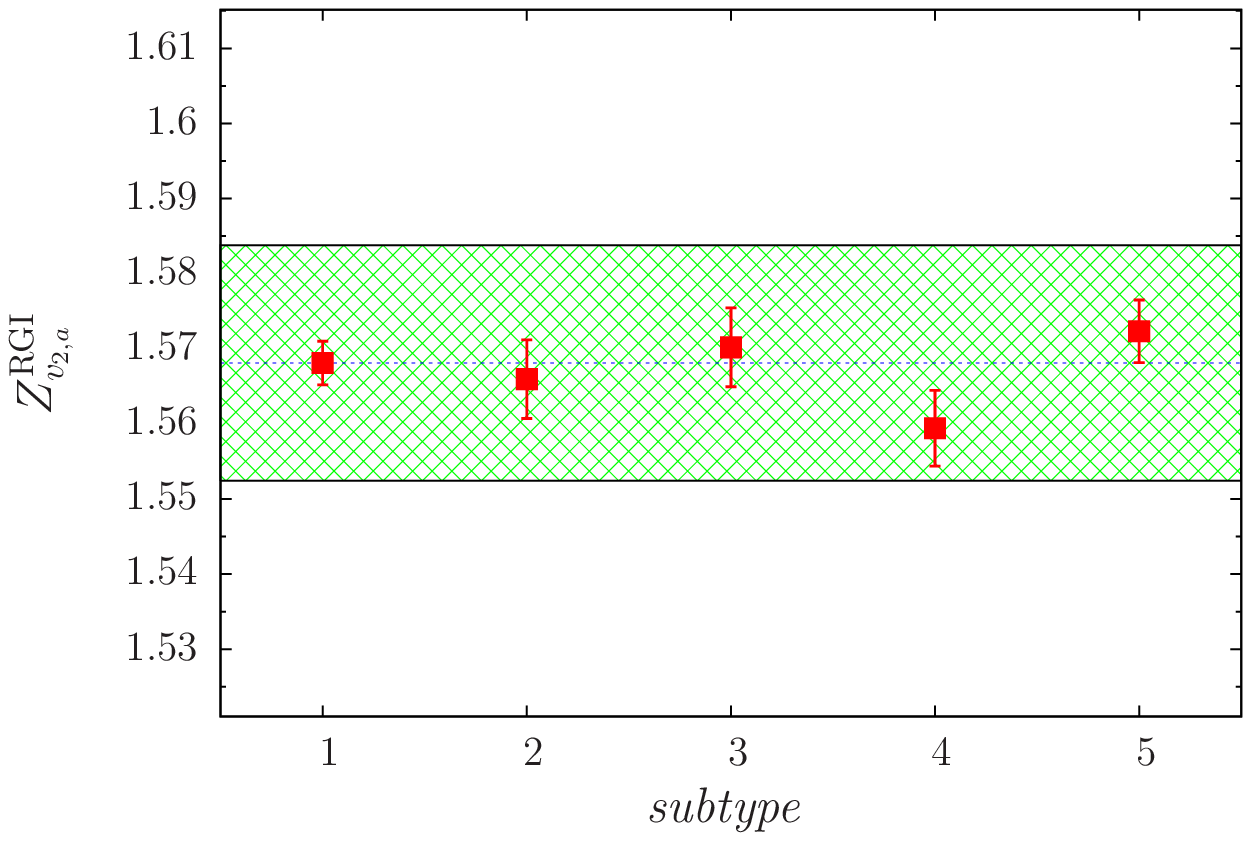}
     \end{tabular}
  \end{center}\vspace{-0.5cm}
  \caption[]{$Z^{\rm RGI}$ of selected operators at  $\beta=5.40$ as a function of the subtraction type
({\it subtype}):
 {\bf 1:} complete subtraction (\ref{Zsuball}) with $g_\star=g_B$,
 {\bf 2:}  ({\bf s}) with $g_\star=g_B$, {\bf 3:}  ({\bf m}) with $g_\star=g_B$,
 {\bf 4:}  ({\bf s}) with $g_\star=g$, {\bf 5:}  ({\bf m}) with $g_\star=g$.
The horizontal borders of the shaded area show a $1\%$ deviation from case {\bf 1}.}
  \label{fig:ZSsubtype}
\end{figure}

In Fig.~\ref{fig:ZSsubtype} 
we  present the $Z^{\rm RGI}$ for operators $\mathcal{O}^S$,  $\mathcal{O}^V$,  $\mathcal{O}^T$
and  $\mathcal{O}^{v_{2,a}}$ for the different subtraction types
using the fit ansatz (\ref{struc1}) 
with all $c_i \neq 0$, i.e., we include $a^2$, $a^4$ and $a^6$ terms. 
From the discussion in Section~\ref{sec:PTSUB22} we expect that the 
resulting differences vary from operator to operator 
(cf.\ Fig.~\ref{fig:ZSZT}).

{}From Fig.~\ref{fig:ZSsubtype} we observe that the complete one-loop subtraction ({\bf 1})
and the subtraction ({\bf 2}) agree within $1\,\%$. This is not unexpected because
the subtraction schemes are similar and the gauge couplings coincide.
The differences in  the results for ({\bf 2}) and ({\bf 3}) can be used as
an indication for a systematic 
uncertainty in the determination of $Z^{\rm RGI}$
based on the schemes ({\bf s,m}). We observe that  both subtraction approaches
are numerically almost equivalent.
Choices ({\bf 4}) and  ({\bf 5}) lead to $Z^{\rm RGI}$ factors which are partly
outside  the $1\,\%$ deviation. 
Generally, we recognize that all subtraction
procedures for both bare and boosted couplings produce fit results within a reasonable error
band width.

\begin{figure}[!thb]
  \begin{center}
     \begin{tabular}{lcr}
        \includegraphics[scale=0.56,clip=true]
         {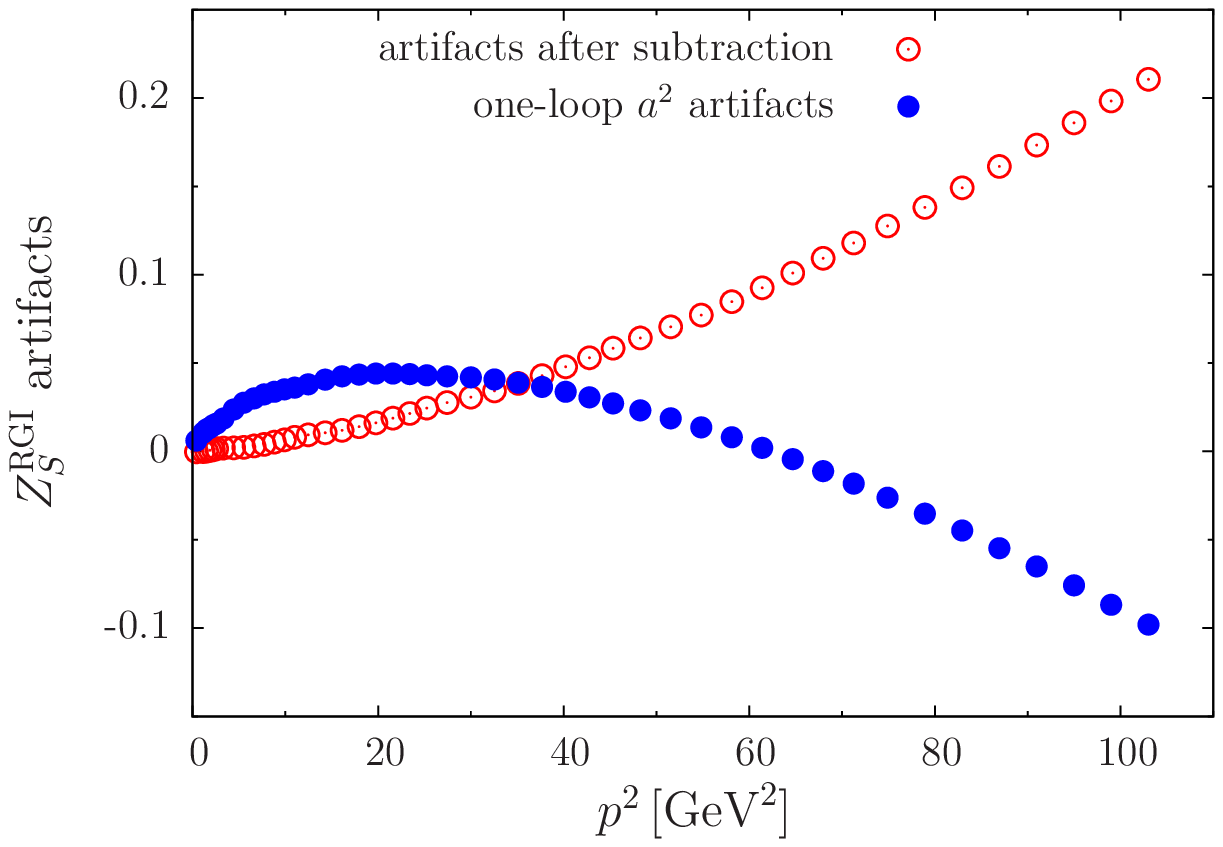}
&&
       \includegraphics[scale=0.56,clip=true]
         {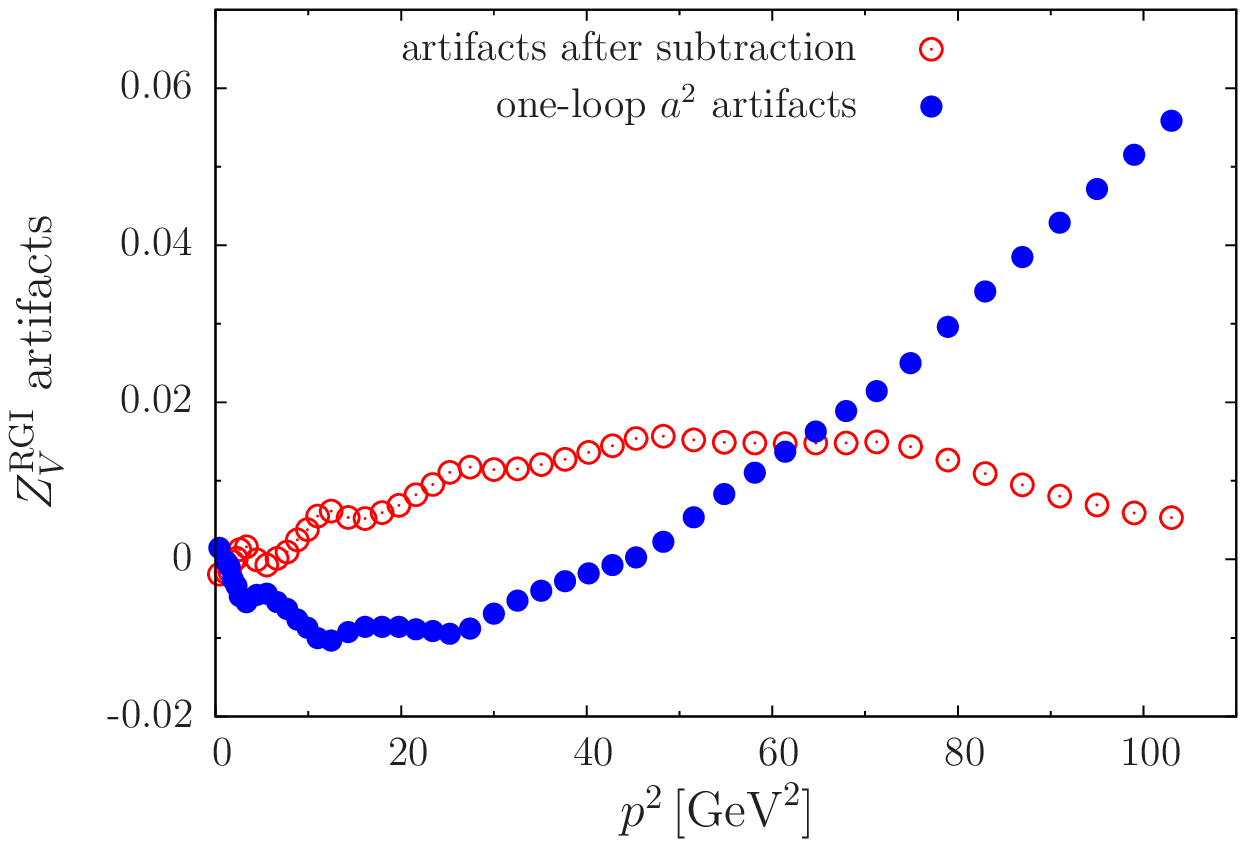}
\\
        \includegraphics[scale=0.56,clip=true]
         {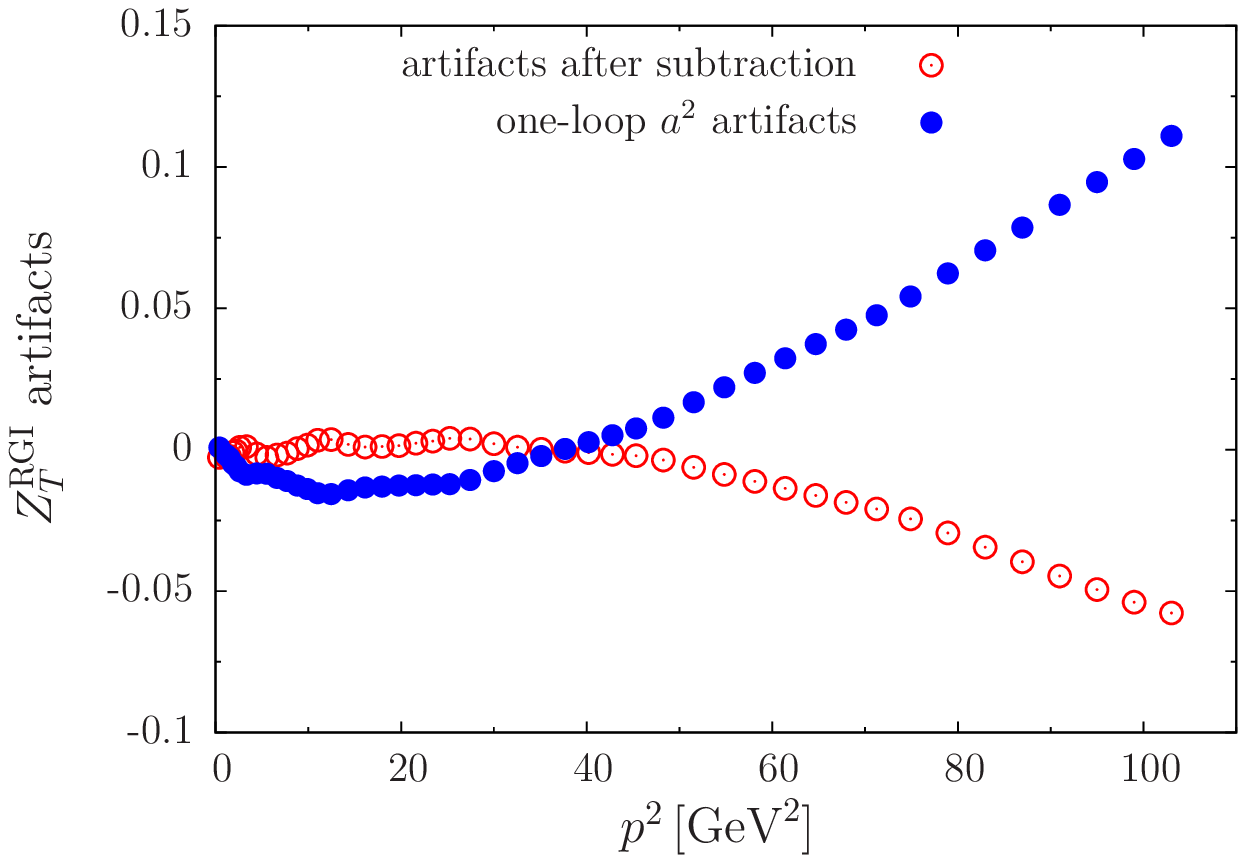}
&&
       \includegraphics[scale=0.56,clip=true]
         {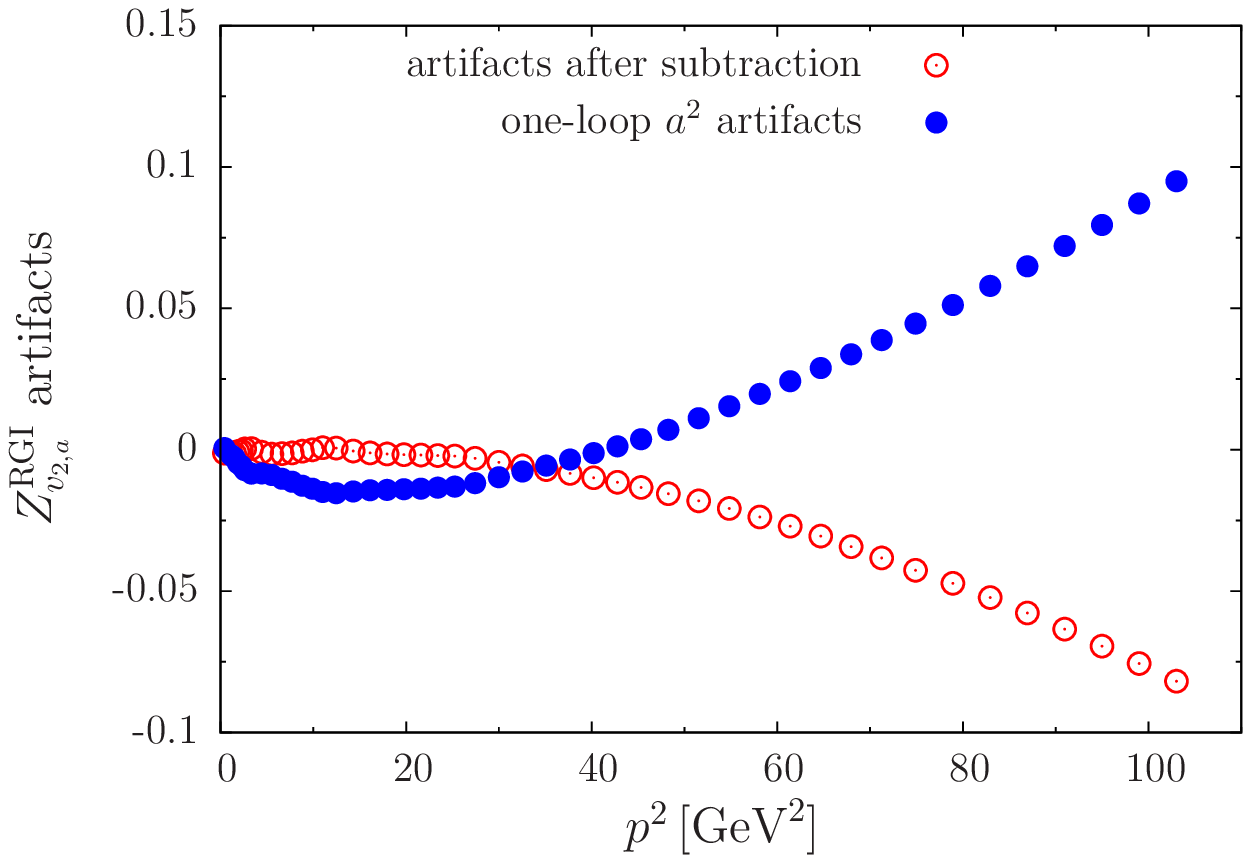}
     \end{tabular}
  \end{center}\vspace{-0.5cm}
  \caption[]{Lattice artifacts for $Z^{\rm RGI}$ of selected operators for  $\beta=5.40$ as a function of 
$p^2$ choosing $g_\star=g_B$. The blue filled circles are the corresponding $g^2 a^2$ correction terms,
the red open circles are the fit results for (\ref{struc1}) setting $Z^{\rm RGI}(a)=0$.}
  \label{fig:Zartifact}
\end{figure}

In order to test  the effect of subtraction we compare the 
$g^2 a^2$ contributions as given in (\ref{Za2all})
with the remaining lattice artifacts of the Monte Carlo data fitted after 
 subtraction, i.e. the result for (\ref{struc1}) setting $Z^{\rm RGI}(a)=0$.
In Fig.~\ref{fig:Zartifact} 
we show the results for the same selected
operators choosing $g_B$. 
In the small $p^2$ region  the remaining
lattice artifacts are significantly smaller than the one-loop
$a^2$ terms (operators $\mathcal{O}^S$, $\mathcal{O}^T$ and $\mathcal{O}^{v_{2,a}}$).
In case of already small one-loop $a^2$ artifacts (operator $\mathcal{O}^V$) the
final artifacts remain small. This behavior strongly suggests to
subtract the one-loop $a^2$ terms before applying the fit procedure.

Since the boosted coupling $g_B$ is
assumed to remove large lattice artifacts due to tadpole contributions in the perturbative series,
we will use $g_B$  in the following.
In addition, we restrict ourselves to
subtraction type ({\bf s}),  which is closest in spirit 
to the complete one-loop subtraction studied in~\cite{Gockeler:2010yr}
(leading approximately to a plateau in the $Z^{\rm RGI}$ as a function of $p^2$).

\subsection{Dependence on hypercubic invariants}

\begin{figure}[!htb]
  \begin{center}
     \begin{tabular}{lcr}
        \includegraphics[scale=0.56,clip=true]
         {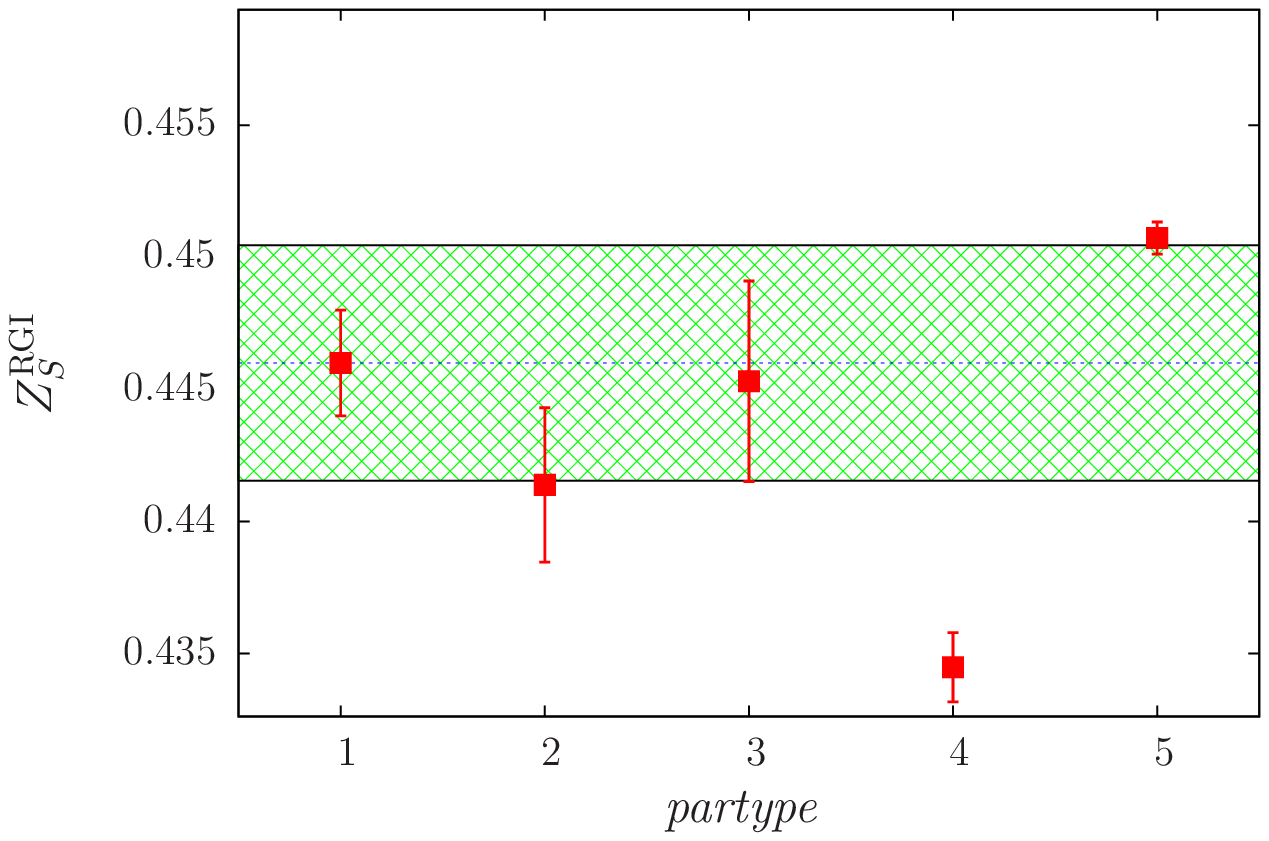}
&&
       \includegraphics[scale=0.56,clip=true]
         {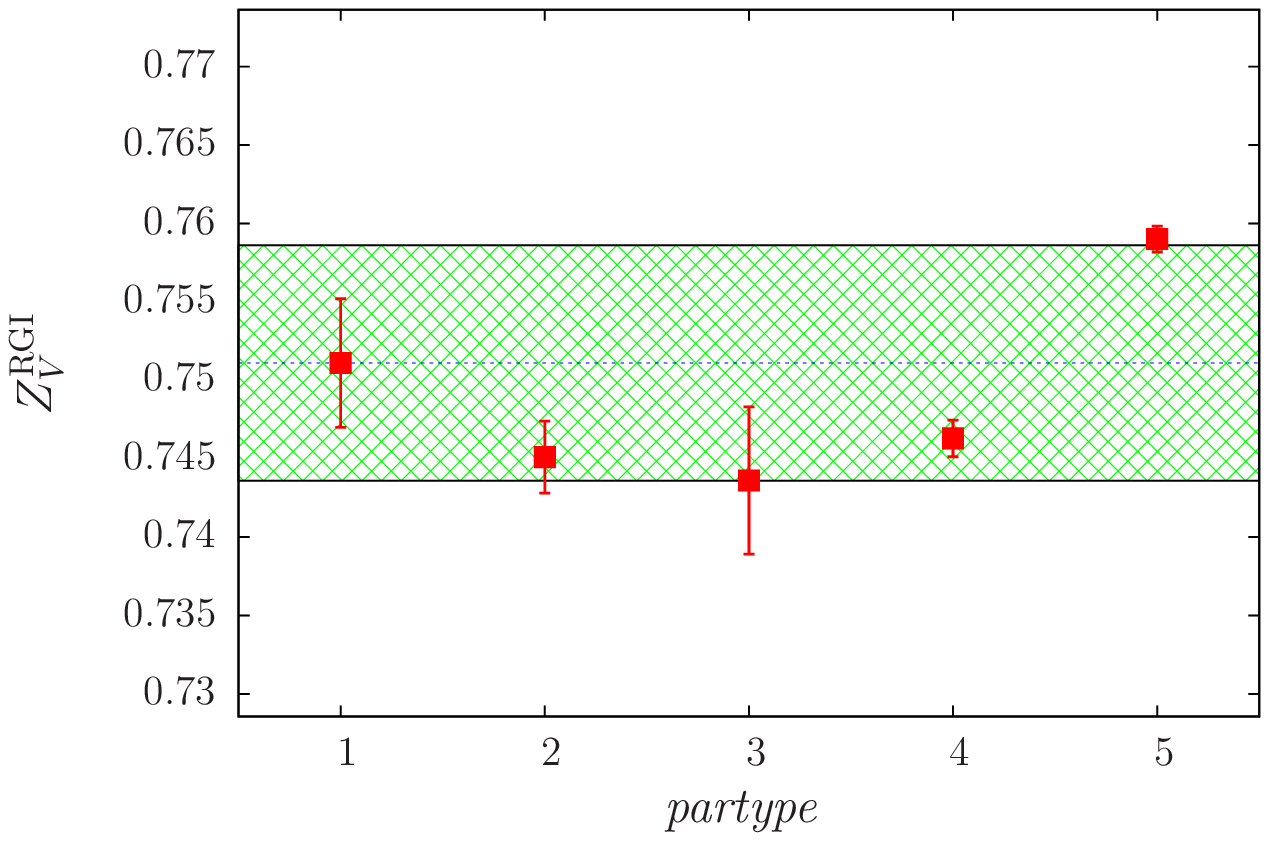}
\\
        \includegraphics[scale=0.56,clip=true]
         {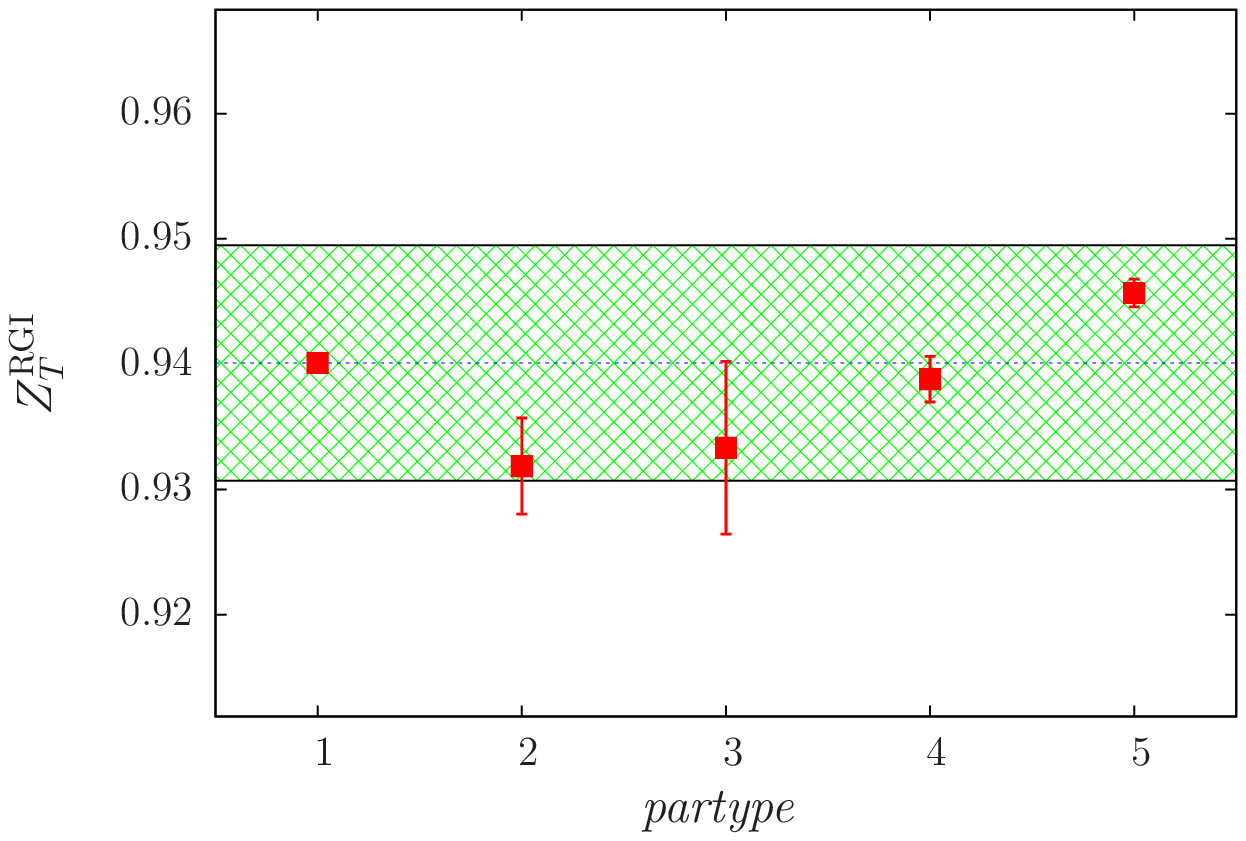}
&&
       \includegraphics[scale=0.56,clip=true]
         {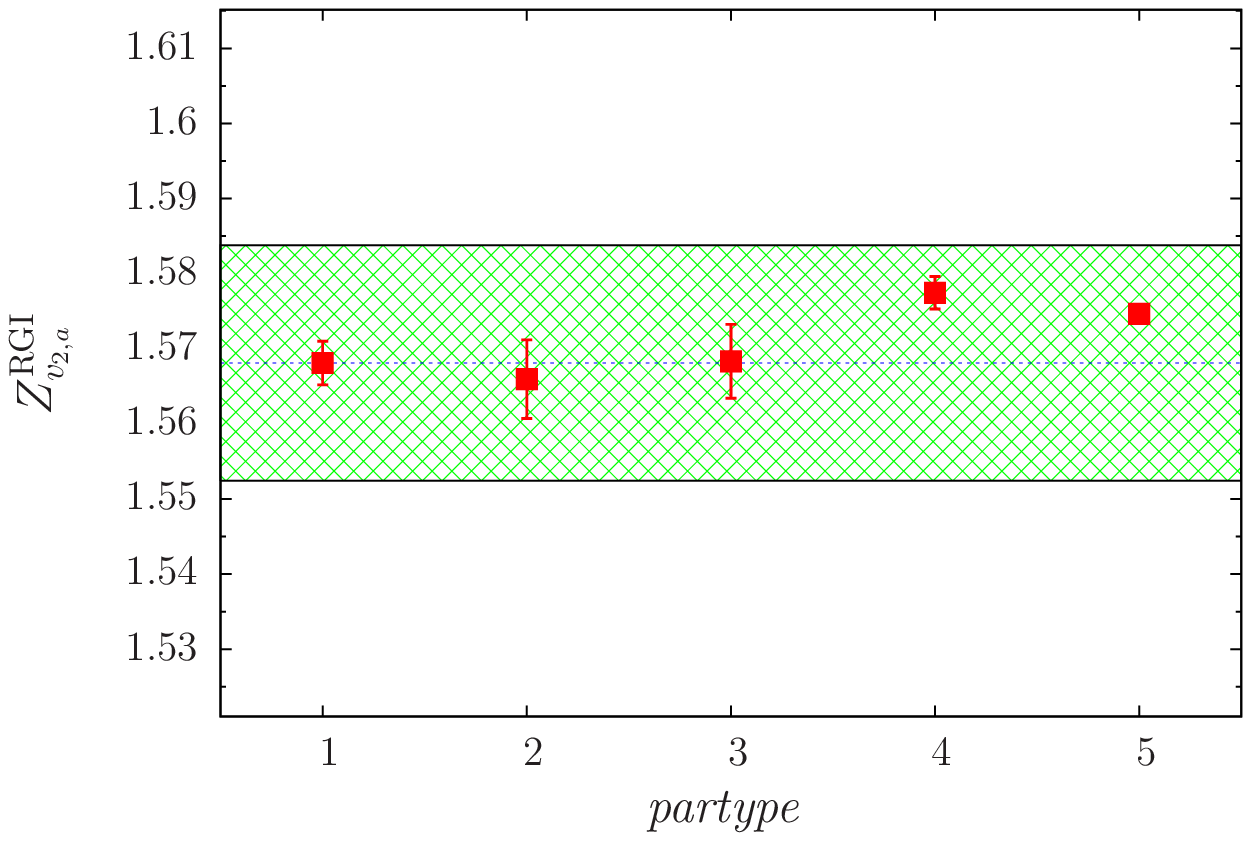}
     \end{tabular}
  \end{center}\vspace{-0.5cm}
  \caption[]{$Z^{\rm RGI}$ for selected operators at $\beta = 5.40$ as a function of the
parameters included in the fit ansatz (\ref{struc1}). The used
parameter combinations
({\it partype}) are: \\
{\bf 1:} complete one-loop subtraction of lattice artifacts (\ref{Zsuball})
{\bf 2:} all $c_i$, {\bf 3:}  $(c_1,c_4,c_6)$ - $O(4)$ invariant, 
{\bf 4:} $(c_1,c_2,c_3,c_4,c_5)$ - ($a^2,a^4$)- hypercubic invariants,
{\bf 5:} $(c_4,c_5,c_6,c_7,c_8)$ - ($a^4,a^6$)- hypercubic invariants.
The horizontal borders of the shaded area show a $1\%$ deviation from case {\bf 1}.}
  \label{fig:ZSpartype}
\end{figure}

Now we discuss the dependence on the hypercubic invariants  included
in the fit ansatz (\ref{struc1}).
 The goal is to 
select a reasonable set of parameters to parametrize the remaining 
lattice artifacts.
Figure~\ref{fig:ZSpartype} 
shows the fit results for some $Z^{\rm RGI}$ utilizing
different parameter sets $\{c_k\}$. 
We use the subtraction type 
({\bf s}) with $g_\star=g_B$. In that case the results from 
the complete one-loop subtraction ({\bf 1}) serve as  reference values.

Generally, we recognize that the resulting
RGI renormalization factors do not vary  significantly.
Most fit results for $Z^{\rm RGI}$ are located in a $1\, \%$
deviation band around the corresponding complete subtraction
results ({\bf 1}). In addition, parametrizations 
({\bf 2}) and ({\bf 3}) give almost identical fit results.
This reflects, of course, the fact that our momenta are very
close to the diagonal in the Brillouin zone.
These restricted momentum sets might be the reason that even
'incomplete' hypercubic invariant sets ({\bf 4, 5}) can be
used to obtain reasonable fits. 
For the final results we use the fit with all $c_i \neq 0$
which would be natural in the case of more
off-diagonal momenta.

\begin{figure}[!htb]
  \begin{center}
     \begin{tabular}{lcr}
        \includegraphics[scale=0.56,clip=true]
         {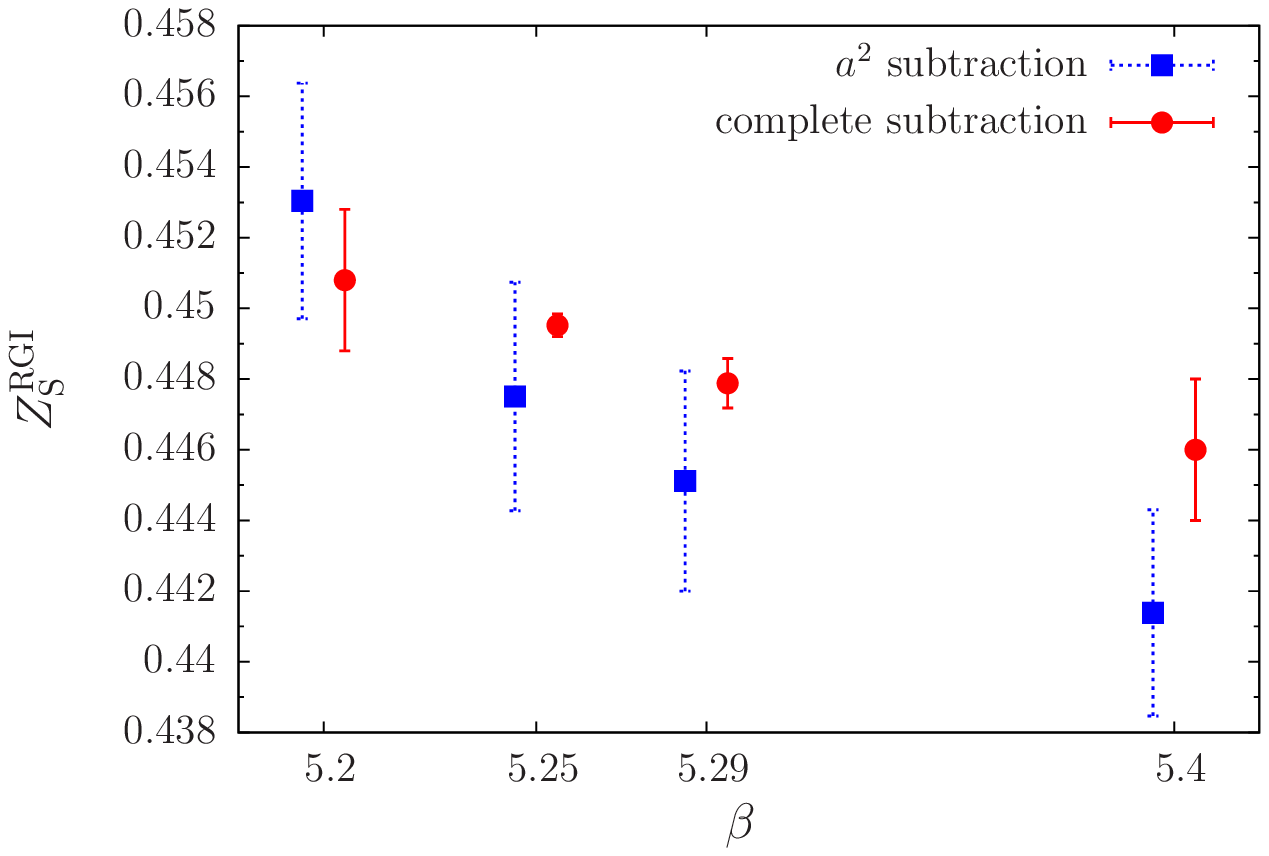}
&&
       \includegraphics[scale=0.56,clip=true]
         {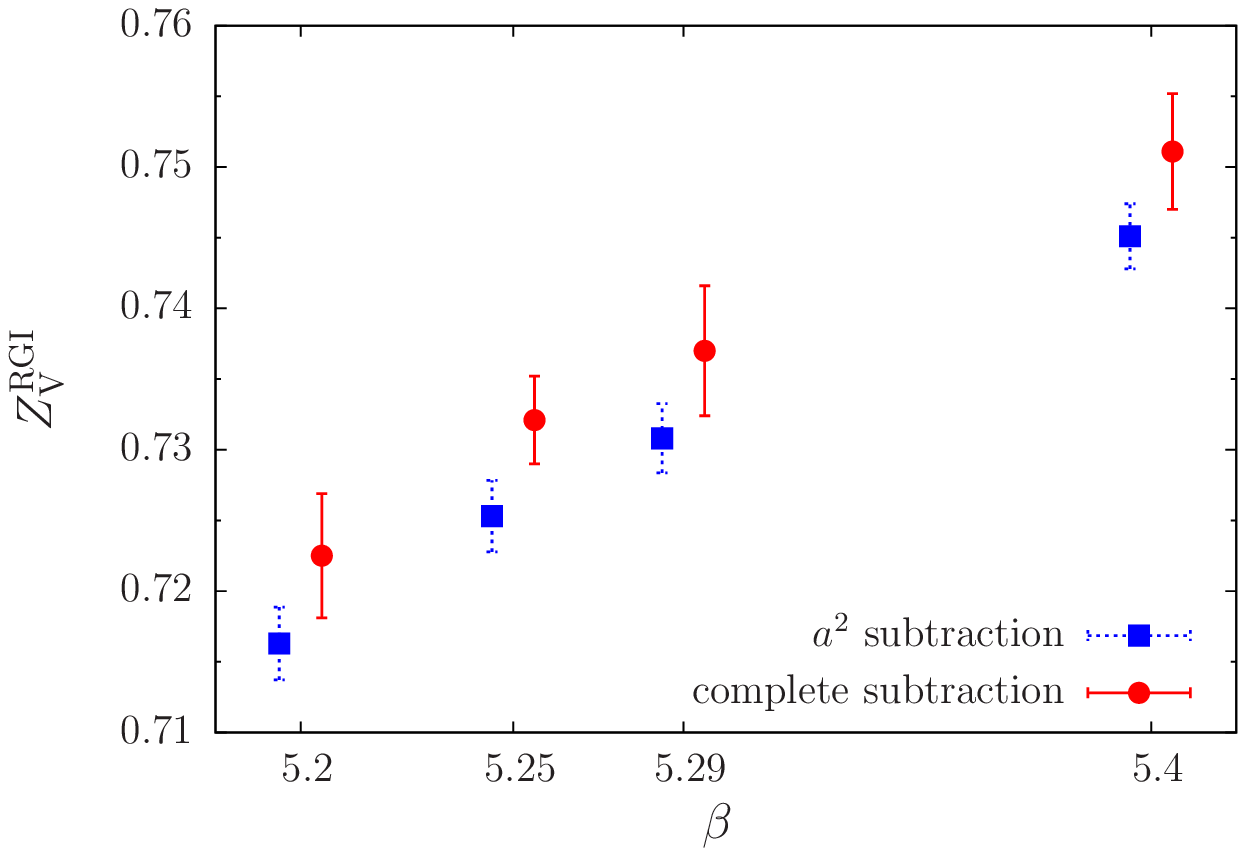}
     \end{tabular}
  \end{center}\vspace{-0.5cm}
  \caption[]{$Z_S^{\rm RGI}$  (left) and $Z_V^{\rm RGI}$ (right) at $r_0\,\Lambda_{\rm \overline{MS}}=0.700$ 
as a function of $\beta$ using
all $c_i$ compared to the complete one-loop 
subtraction.}
  \label{fig:pSV}
\end{figure}

\begin{figure}[!htb]
  \begin{center}
     \begin{tabular}{lcr}
        \includegraphics[scale=0.56,clip=true]
         {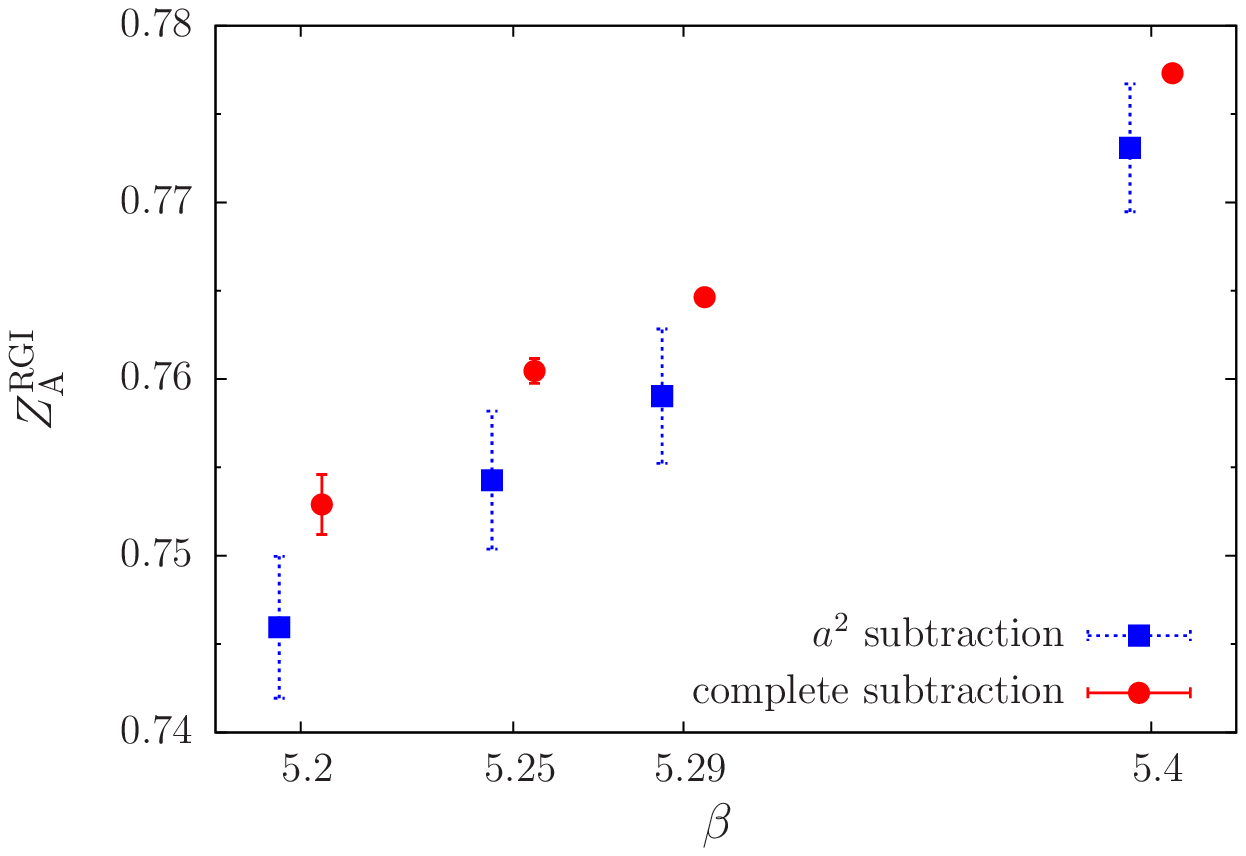}
&&
       \includegraphics[scale=0.56,clip=true]
         {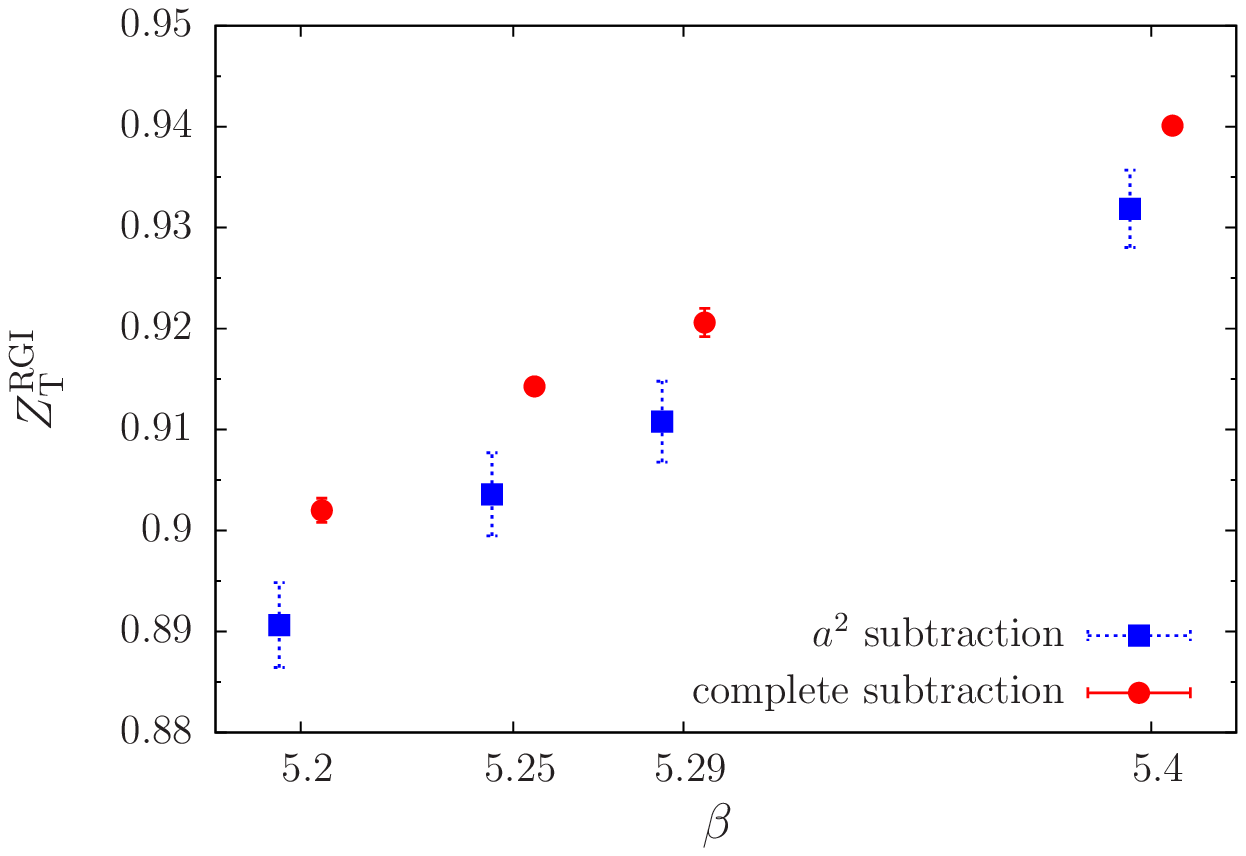}
     \end{tabular}
  \end{center}\vspace{-0.5cm}
  \caption[]{The same as in Fig.~\ref{fig:pSV} for $Z_A^{\rm RGI}$  (left) and $Z_T^{\rm RGI}$ (right).}
  \label{fig:pAT}
\end{figure}

\begin{figure}[!htb]
  \begin{center}
     \begin{tabular}{lcr}
        \includegraphics[scale=0.56,clip=true]
         {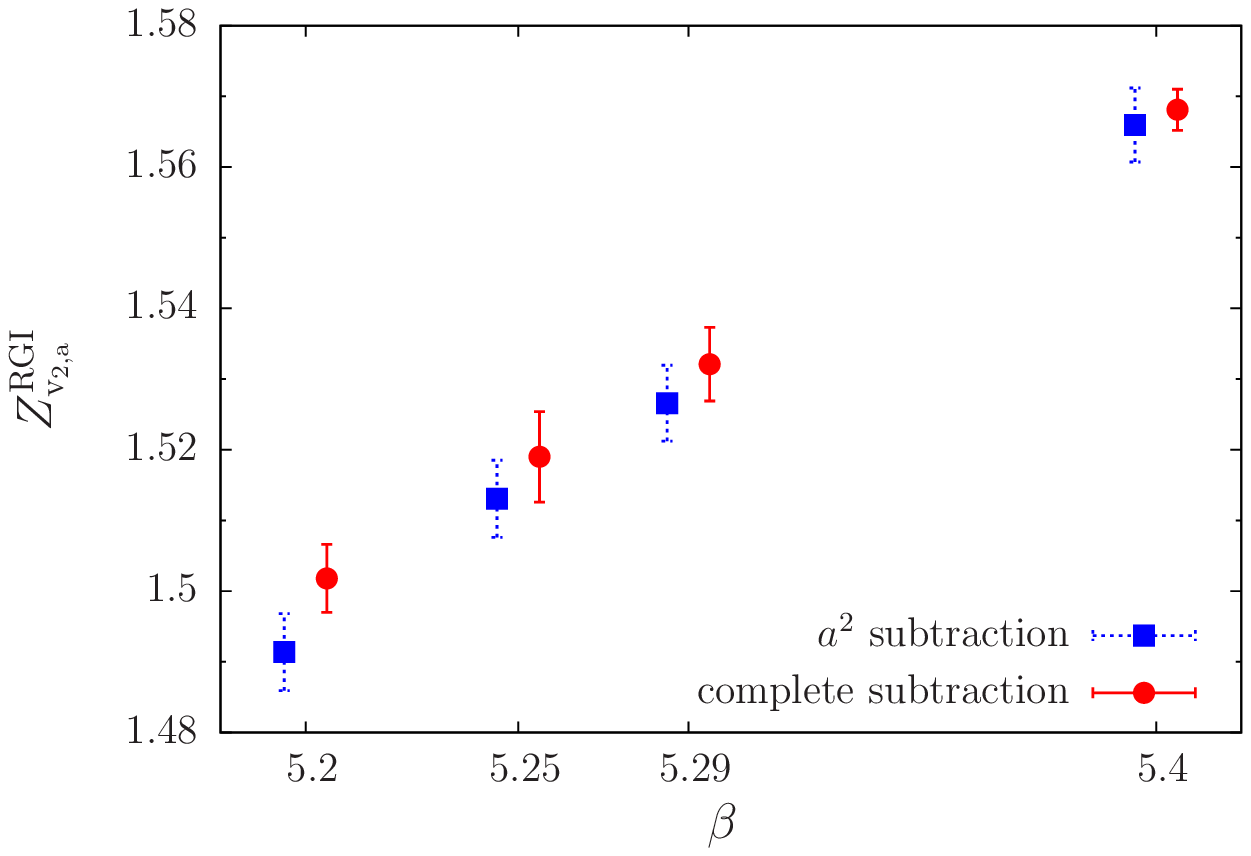}
&&
       \includegraphics[scale=0.56,clip=true]
         {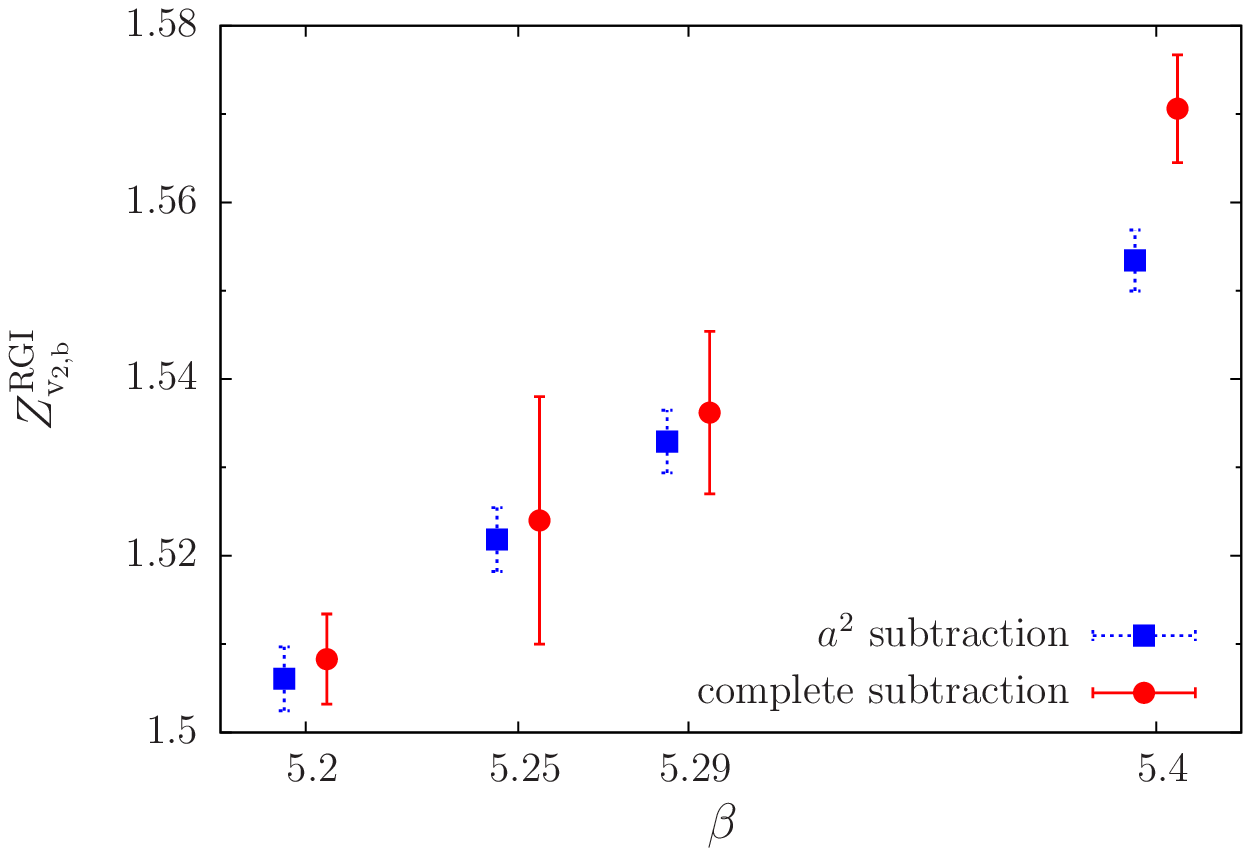}
     \end{tabular}
  \end{center}\vspace{-0.5cm}
  \caption[]{The same as in Fig.~\ref{fig:pSV} for $Z_{v_{2,a}}^{\rm RGI}$  (left) and $Z_{v_{2,b}}^{\rm RGI}$ (right).}
  \label{fig:pvavb}
\end{figure}

In Figs.~\ref{fig:pSV}, \ref{fig:pAT} and \ref{fig:pvavb} we show the results for all 
operators using the parameter sets with all $c_i$ 
compared to the results obtained by the subtraction scheme based on (\ref{Zsuball}). 

\section{Results for local and one-link operators and conclusions}

As a result of the preceding discussions we use 
subtraction type
({\bf s}) (eq. (\ref{eq:pts11})) with boosted coupling $g_B$
and
the fitting formula (\ref{struc1}) with all $c_i$ and $b_1$ coefficients
to determine the $Z^{\rm RGI}$. 
The final renormalization factors are collected in Table~\ref{tabZRGIm}
\renewcommand*\arraystretch{1.5}
\begin{table}[!htb]
\begin{center}
  \begin{tabular}{||c|c|c|c|c|c||}
\hline
Op. & $r_0\,\Lambda_{\rm \overline{MS}}$ & $Z^{\rm RGI}\vert_{\beta=5.20}$ & $Z^{\rm RGI}\vert_{\beta=5.25}$ &  $Z^{\rm RGI}\vert_{\beta=5.29}$ & $Z^{\rm RGI}\vert_{\beta=5.40}$ \\ 
\hline
$\mathcal{O}^S $&$ 0.700 $&$  0.4530(34) $&$  0.4475(33) $&$  0.4451(32) $&$  0.4414(30)$\\
$  $&$ 0.789 $&$  0.4717(44) $&$  0.4661(65) $&$  0.4632(54) $&$  0.4585(27)$\\
\hline
$\mathcal{O}^V $&$ 0.700 $&$  0.7163(26) $&$  0.7253(26) $&$  0.7308(25) $&$  0.7451(24)$\\
$  $&$ 0.789 $&$  0.7238(72) $&$  0.7319(94) $&$  0.7365(99) $&$  0.7519(50)$\\
\hline
$\mathcal{O}^A $&$ 0.700 $&$  0.7460(41) $&$  0.7543(40) $&$  0.7590(39) $&$  0.7731(37)$\\
$  $&$ 0.789 $&$  0.7585(46) $&$  0.7634(77) $&$  0.7666(81) $&$  0.7805(30)$\\
\hline
$\mathcal{O}^T $&$ 0.700 $&$  0.8906(43) $&$  0.9036(42) $&$  0.9108(41) $&$  0.9319(39)$\\
$  $&$ 0.789 $&$  0.8946(85) $&$  0.9041(111) $&$  0.9075(120) $&$  0.9316(49)$\\
\hline
$\mathcal{O}^{v_{2,a}} $&$ 0.700 $&$  1.4914(55) $&$  1.5131(55) $&$  1.5266(54) $&$  1.5660(53)$\\
$  $&$ 0.789 $&$  1.4635(108) $&$  1.4776(112) $&$  1.4926(90) $&$  1.5397(58)$\\
\hline
$\mathcal{O}^{v_{2,b}} $&$ 0.700 $&$  1.5061(37) $&$  1.5218(37) $&$  1.5329(36) $&$  1.5534(35)$\\
$  $&$ 0.789 $&$  1.4601(151) $&$  1.4727(206) $&$  1.4863(165) $&$  1.5115(140)$\\
\hline
\end{tabular}
 \end{center}
  \caption{$Z^{\rm RGI}$ values using the subtraction ({\bf s}) with $g_B$.}
\label{tabZRGIm}
\end{table}
using the two different $r_0\,\Lambda_{\rm \overline{MS}}$ values $0.700$
and $0.789$.
This shows  the influence of the choice of 
$r_0\,\Lambda_{\rm \overline{MS}}$ (depending on the anomalous dimension
of the operator). For the investigated operators and
$\beta$ values we found for the relative differences
of the $Z^{\rm RGI}$ 
\begin{equation}
 \delta Z^{\rm RGI} = \Bigg|\frac{Z^{\rm RGI}_{r_0\,\Lambda_{\rm \overline{MS}}=0.700}-
Z^{\rm RGI}_{r_0\,\Lambda_{\rm \overline{MS}}=0.789}} 
{Z^{\rm RGI}_{r_0\,\Lambda_{\rm \overline{MS}}=0.700}}\Bigg| \lesssim 0.04\,.
\end{equation}

For comparison we collect in Table~\ref{tabZRGIold}
 the values for
$Z^{\rm RGI}$ computed by means of fits with the ansatz (\ref{RGI4n})
to data where a complete one-loop
subtraction of lattice artifacts (according to (\ref{Zsuball}) 
with $g_{\star} = g_{\rm B}$) has been performed.
Note that here the errors are determined from the variation of the subtracted
data between the scales $\mu^2 = 10, \, 20, \, 30 \, {\rm GeV}^2$ 
\cite{Gockeler:2010yr}. 
\begin{table}[!htb]
\begin{center}
  \begin{tabular}{||c|c|c|c|c|c||}
\hline
Op. & $r_0\,\Lambda_{\rm \overline{MS}}$ & $Z^{\rm RGI}\vert_{\beta=5.20}$ & $Z^{\rm RGI}\vert_{\beta=5.25}$ &  $Z^{\rm RGI}\vert_{\beta=5.29}$ & $Z^{\rm RGI}\vert_{\beta=5.40}$ \\ 
\hline
$\mathcal{O}^S $&$ 0.700 $& $0.4508(20)$  & $0.44952(32)$ & $0.44788(70)$ & $0.4460(20)$    \\
$  $&$ 0.789 $& $0.4620(85)$  & $0.4603(60)$  & $0.4585(61)$  & $0.4560(48)$    \\
\hline
$\mathcal{O}^V $&$ 0.700 $& $0.7225(44)$  & $0.7321(31)$  & $0.7370(46)$  & $0.7511(41)$    \\
$  $&$ 0.789 $& $0.7219(53)$  & $0.7316(41)$  & $0.7364(55)$  & $0.7506(50)$    \\
\hline
$\mathcal{O}^A $&$ 0.700 $& $0.7529(17)$  & $0.76046(70)$ & $0.76463(33)$ & $0.77731(20)$   \\
$  $&$ 0.789 $& $0.7530(14)$  & $0.76054(48)$ & $0.7647(14)$  & $0.7774(10)$    \\
\hline
$\mathcal{O}^T $&$ 0.700 $& $0.9020(12)$  & $0.91427(24)$ & $0.9206(14)$  & $0.94009(69)$   \\
$  $&$ 0.789 $& $0.8948(40)$  & $0.9072(32)$  & $0.9137(48)$  & $0.9333(38)$    \\
\hline
$\mathcal{O}^{v_{2,a}} $&$ 0.700 $& $1.5018(48)$  & $1.5190(64)$  & $1.5321(52)$  & $1.5681(29)$    \\
$  $&$ 0.789 $& $1.473(18)$   & $1.490(14)$   & $1.504(12)$   & $1.540(14)$     \\
\hline
$\mathcal{O}^{v_{2,b}} $&$ 0.700 $& $1.5083(51)$  & $1.524(14)$   & $1.5362(92)$  & $1.5706(61)$    \\
$  $&$ 0.789 $& $1.480(15)$   & $1.497(28)$   & $1.509(23)$   & $1.5436(69)$    \\
\hline
\end{tabular}
 \end{center}
  \caption{$Z^{\rm RGI}$ using a complete
           one-loop subtraction of lattice artifacts.}
\label{tabZRGIold}
\end{table}
The reported renormalization factors are 
calculated for the values $r_0/a$ given at the end of Section~\ref{sec:PTSUB2}
and, therefore, differ from those given in~\cite{Gockeler:2010yr}.
The $Z$ factors of the local operators in both tables agree  within $1\,\%$.
The  $Z$ factors of the one-link operators differ at most by $2\,\%$.

Let us compare our results in Table~\ref{tabZRGIold}  for the local vector current with $Z_V^{\rm RGI}$ 
obtained from an 
analysis of the proton electromagnetic form factor~\cite{Collins:2011mk} following~\cite{Bakeyev:2003ff}, 
which are listed in Table~\ref{tabZV}. 
The numbers agree within less than $1\,\%$ with the numbers in 
Table~\ref{tabZRGIold} ($r_0\,\Lambda_{\rm \overline{MS}}=0.700$), supporting the complete one-loop 
subtraction as our reference point. 
\renewcommand*\arraystretch{1.5}
\begin{table}[!htb]
\begin{center}
\vspace{0.5cm}
  \begin{tabular}{||c|c|c|c||}
\hline
$Z^{\rm RGI}\vert_{\beta=5.20}$ & $Z^{\rm RGI}\vert_{\beta=5.25}$ &  $Z^{\rm RGI}\vert_{\beta=5.29}$ & $Z^{\rm RGI}\vert_{\beta=5.40}$ \\ 
\hline
 $0.7296(4)$  & $0.7355(3)$ & $0.7401(2)$ & $0.7521(3)$    \\
\hline
\end{tabular}
 \end{center}
  \caption{$Z^{\rm RGI}$ values for operator $V$ from the proton electromagnetic form factor analysis.}
\label{tabZV}
\end{table}

From the present investigation we conclude:
The alternatively
proposed 'reduced' subtraction algorithm can be used for the
determination of the renormalization factors if the complete
subtraction method is not available.
Possible applications could be $Z$ factors for $N_f = 2+1$ 
calculations with more complicated fermionic and gauge actions
where one-loop results to order $a^2$ are available
(for the fermionic SLiNC action with improved 
Symanzik gauge action see Ref.~\cite{Skouroupathis:2010zz}).

In this study we have analyzed data sets with momenta close to the diagonal
of the Brillouin zone. 
The one-loop $a^2$ contributions to the $Z$ factors are completely
general and can be used for arbitrary 
(also non-diagonal) momentum sets.
Our ansatz (\ref{struc1}) allows 
to take into account the remaining artifacts after  subtracting
these one-loop $a^2$ terms.
To get reasonable
fit results the ratio (number of data points)/(number of fit parameters)
has to be sufficiently large.

As we pointed out the subtraction type is not unique. With ({\bf s})
and ({\bf m}) we tested two different types. The resulting fits
do not give a clear preference for one of these. Even the additional choice for  the 
coupling ($g_\star = g$ or $g_\star = g_B$) does not lead to  significantly 
different results. Therefore, our final choice ({\bf s}) (eq. (\ref{eq:pts11}) with  $g_\star = g_B$)
was supported by 'external' arguments: the improved behavior of the boosted
perturbative series and the results obtained by complete one-loop subtraction~\cite{Gockeler:2010yr}.

We have shown that already the one-loop $a^2$ subtraction improves the behavior
of the $Z$ factors significantly: In the small $p^2$ region the contributions
of the remaining lattice artifacts are  smaller
than the corresponding one-loop $a^2$ terms.
As mentioned above, the accuracy to determine the $Z$ factors is already 
at the $1\,\%$ level for local operators and at the $2\,\%$ level for operators
with one covariant derivative compared to the complete one-loop subtraction
of lattice artifacts. 
Additional systematic uncertainties are due
to the choice of the $r_0\,\Lambda_{\rm \overline{MS}}$ and $r_0/a$.

\renewcommand{\thesection}{\Alpha{section}}
\renewcommand{\theequation}{A.\arabic{equation}}
\setcounter{equation}{0}
\setcounter{section}{0}
\section*{Appendix}

In this Appendix we show that the definition (\ref{ZRinv}) leads to 
renormalization factors which are invariant under the hypercubic 
group $H(4)$.

We consider a multiplet of local quark-antiquark operators $\mathcal O_i (x)$ 
($i=1,2,\ldots,d$) in position space which transform according to 
\begin{equation} 
\mathcal O_i (x) \to S_{ij} (R) \, \mathcal O_j (R^{-1}x)
\end{equation} 
when
\begin{equation} 
\psi (x) \to D(R)\, \psi (R^{-1}x) \; , \;
\bar{\psi} (x) \to \bar{\psi} (R^{-1}x) \, D(R)^\dagger
\end{equation} 
for all $N=384$ elements $R$ of $H(4)$. Here $D(R)$ denotes the (unitary) 
spinor representation of $H(4)$ (or $O(4)$):
\begin{equation} 
D(R)^\dagger \gamma_\mu D(R) = R_{\mu \nu} \gamma_\nu \,.
\end{equation}
We assume that the operators $\mathcal O_i (x)$ have been chosen such that
the $d \times d$-matrices $S(R)$ form a unitary irreducible representation 
of $H(4)$. 

Denoting the unrenormalized vertex function at external momentum $p$ of the operator $\mathcal O_i$
by $\Gamma_i (p)$ we have
\begin{equation} 
\Gamma_i (p) = \sum_{j=1}^d S_{ij} (R)\, D(R) \,\Gamma_j (R^{-1}p)\, D(R)^\dagger
\end{equation}
for all $R \in H(4)$, and analogously for the corresponding Born term 
$\Gamma_i^{\mathrm {Born}} (p)$. Consequently we get
\begin{equation} 
\sum_{i=1}^d \mbox{tr} \left[ \Gamma_i (p) \Gamma_i (p)^\dagger \right] =
\sum_{i=1}^d \mbox{tr} \left[ \Gamma_i (Rp) \Gamma_i (Rp)^\dagger \right]\,.
\end{equation}
Using the orthogonality relations for the matrix 
elements of irreducible representations one finds in addition
\begin{equation} 
\sum_R \mbox{tr} \left[ \Gamma_i (Rp) \Gamma_j (Rp)^\dagger \right] =
\frac{1}{d} \delta_{ij} \sum_{k=1}^d \sum_R \mbox{tr} 
   \left[ \Gamma_k (Rp) \Gamma_k (Rp)^\dagger \right] \,,
\end{equation}
where the sum extends over all $R \in H(4)$. The same relations
hold when one of the vertex functions or both are replaced by 
the corresponding Born terms, e.g.,
\begin{equation} 
\sum_{i=1}^d \mbox{tr} 
  \left[ \Gamma_i (p) \Gamma_i^{\mathrm {Born}} (p)^\dagger \right] =
\sum_{i=1}^d \mbox{tr} 
  \left[ \Gamma_i (Rp) \Gamma_i^{\mathrm {Born}} (Rp)^\dagger \right] \,.
\end{equation}

Therefore the renormalization condition
\begin{equation} 
Z^{-1} Z_q  = \frac{ 
  \sum_{i=1}^d \mbox{tr} \left[ \Gamma_i (p) \, 
                    \Gamma_i^{\mathrm {Born}} (p)^\dagger \right]}
 {\sum_{j=1}^d \mbox{tr} \left[ \Gamma_j^{\mathrm {Born}} (p) \,
                    \Gamma_j^{\mathrm {Born}} (p)^\dagger \right]} 
\end{equation}
or, equivalently, 
\begin{equation} 
Z^{-1} Z_q \delta_{ij} = \frac{d}{N} \frac{ \sum_R 
   \mbox{tr} \left[ \Gamma_i (Rp) \, 
                    \Gamma_j^{\mathrm {Born}} (Rp)^\dagger \right]}
 {\sum_{k=1}^d \mbox{tr} \left[ \Gamma_k^{\mathrm {Born}} (p) \,
                    \Gamma_k^{\mathrm {Born}} (p)^\dagger \right]} 
\end{equation}
respects the hypercubic symmetry, i.e., writing more 
precisely $Z=Z(p)$ we have $Z(Rp) = Z(p)$ for all $R \in H(4)$, 
and all lattice artefacts in $Z$ must be invariant under the hypercubic 
group. Of course, here it has been assumed that $Z_q(Rp) = Z_q(p)$,
as is the case for our definition (\ref{defzq}) of $Z_q$.

\section*{Acknowledgements}
\vspace{-1mm}
This work has been supported in part by the DFG under contract 
SFB/TRR55 (Hadron Physics from Lattice QCD) and by the EU grant 283286 (HadronPhysics3).
M. Constantinou, M. Costa and H. Panagopoulos acknowledge support from
the Cyprus Research Promotion Foundation under Contract No.
TECHNOLOGY/$\Theta$E$\Pi$I$\Sigma$/ 0311(BE)/16. 
We thank D. Pleiter for providing us the 
current values of $Z_V^{\rm RGI}$
from an analysis of the proton electromagnetic form factor.


\begin{thebibliography}{99}

\bibitem{Capitani:2002mp}
  S.~Capitani,
  Phys.\ Rept.\  {\bf 382} (2003) 113
[\href{http://arxiv.org/abs/hep-lat/0211036}{arXiv:hep-lat/0211036}].

\bibitem{Martinelli:1994ty}
  G.~Martinelli, C.~Pittori, C.~T.~Sachrajda, M.~Testa and A.~Vladikas,
  Nucl.\ Phys.\ B {\bf 445} (1995) 81
[\href{http://arxiv.org/abs/hep-lat/9411010}{arXiv:hep-lat/9411010}].


\bibitem{Arthur:2010ht}
  R.~Arthur and P.~A.~Boyle (RBC and UKQCD Collaborations),
  Phys.\ Rev.\ D {\bf 83} (2011) 114511
[\href{http://arxiv.org/abs/1006.0422}{arXiv:1006.0422[hep-lat]}].

\bibitem{Gockeler:2010yr}
M.~G\"ockeler, R.~Horsley, Y.~Nakamura, H.~Perlt, D.~Pleiter, P.~E.~L.~Rakow, A.~Sch\"afer, 
G.~Schierholz, A.~Schiller, H. St\"uben and J.~M.~Zanotti, (QCDSF/UKQCD Collaboration)
  Phys.\ Rev.\  D {\bf 82} (2010) 114511 
  [Erratum-ibid.\ D {\bf 86} (2012) 099903]
[\href{http://arxiv.org/abs/1003.5756}{arXiv:1003.5756[hep-lat]}].
  
\bibitem{Gockeler:1996mu}
  M.~G\"ockeler, R.~Horsley, E.-M.~Ilgenfritz, H.~Perlt, P.~E.~L.~Rakow, G.~Schierholz and A.~Schiller,
  Phys.\ Rev.\ D {\bf 54} (1996) 5705
[\href{http://arxiv.org/abs/hep-lat/9602029}{arXiv:hep-lat/9602029}].


\bibitem{Constantinou:2009tr}
  M.~Constantinou, V.~Lubicz, H.~Panagopoulos and F.~Stylianou,
  JHEP {\bf 0910} (2009) 064
[\href{http://arxiv.org/abs/0907.0381}{arXiv:0907.0381[hep-lat]}].
 


\bibitem{Larin:1993vu}
  S.~A.~Larin, T.~van Ritbergen and J.~A.~M.~Vermaseren,
  Nucl.\ Phys.\ B {\bf 427} (1994) 41.

\bibitem{Retey:2000nq}
  A.~Retey and J.~A.~M.~Vermaseren,
  Nucl.\ Phys.\ B {\bf 604} (2001) 281
[\href{http://arxiv.org/abs/hep-ph/0007294}{arXiv:hep-ph/0007294}].



\bibitem{Gracey:2006zr}
  J.~A.~Gracey,
  JHEP {\bf 0610} (2006) 040
[\href{http://arxiv.org/abs/hep-ph/0609231}{arXiv:hep-ph/0609231}].
  
  
 
\bibitem{Skouroupathis:2010zz} 
  A.~Skouroupathis and H.~Panagopoulos,
  PoS LATTICE {\bf 2010}, 240 (2010).
 
\bibitem{Alexandrou:2012mt}
  C.~Alexandrou, M.~Constantinou, T.~Korzec, H.~Panagopoulos and F.~Stylianou,
  Phys.\ Rev.\ D {\bf 86} (2012) 014505
[\href{http://arxiv.org/abs/1201.5025}{arXiv:1201.5025[hep-lat]}].



\bibitem{mathematica}
Mathematica, Version 9.0, Wolfram Research, Inc., Champaign, IL (2012)

\bibitem{minuit}
MINUIT, Reference Manual, F.~James, CERN Geneva, Switzerland (1994)

\bibitem{QCDSF:2013}
  QCDSF collaboration, in preparation.

\bibitem{Fritzsch:2012wq}
  P.~Fritzsch, F.~Knechtli, B.~Leder, M.~Marinkovic, S.~Schaefer, R.~Sommer and F.~Virotta (ALPHA Collaboration),
  Nucl.\ Phys.\ B {\bf 865} (2012) 397
[\href{http://arxiv.org/abs/1205.5380}{arXiv:1205.5380[hep-lat]}].


\bibitem{Bali:2012qs}
  G.~S.~Bali, P.~C.~Bruns, S.~Collins, M.~Deka, B.~Gl\"a\ss le, M.~G\"ockeler,
L.~Greil, T.R.~Hemmert, R.~Horsley, J.~Najjar, Y.~Nakamura, A.~Nobile,
D.~Pleiter, P.~E.~L.~Rakow, A.~Sch\"afer, R.~Schiel, G.~Schierholz, A.~Sternbeck
and J.~M.~Zanotti (QCDSF Collaboration),
  Nucl.\ Phys.\ B {\bf 866} (2013) 1
[\href{http://arxiv.org/abs/1206.7034}{arXiv:1206.7034[hep-lat]}].
  


\bibitem{Boucaud:2003dx}
  P.~Boucaud, F.~de Soto, J.~P.~Leroy, A.~Le Yaouanc, J.~Micheli, H.~Moutarde,
O.~Pene and J.~Rodriguez-Quintero,
  Phys.\ Lett.\ B {\bf 575} (2003) 256
[\href{http://arxiv.org/abs/hep-lat/0307026}{arXiv:hep-lat/0307026}].

\bibitem{deSoto:2007ht}
  F.~de Soto and C.~Roiesnel,
  JHEP {\bf 0709} (2007) 007
[\href{http://arxiv.org/abs/0705.3523}{arXiv:0705.3523[hep-lat]}].


\bibitem{Collins:2011mk}
S.~Collins, M.~G\"ockeler, P.~H\"agler, R.~Horsley, Y.~Nakamura, A.~Nobile, D.~Pleiter, P.~E.~L.~Rakow, 
A.~Sch\"afer, G.~Schierholz, W.~Schroers, H.~St\"uben, F.~Winter and J.~M.~Zanotti (QCDSF/UKQCD Collaboration),
  Phys.\ Rev.\ D {\bf 84} (2011) 074507
[\href{http://arxiv.org/abs/1106.3580}{arXiv:1106.3580[hep-lat]}].



\bibitem{Bakeyev:2003ff}
  T.~Bakeyev, M.~G\"ockeler, R.~Horsley, D.~Pleiter, P.~E.~L.~Rakow, G.~Schierholz and 
H.~St\"uben (QCDSF/UKQCD Collaboration),
  Phys.\ Lett.\ B {\bf 580} (2004) 197
[\href{http://arxiv.org/abs/hep-lat/0305014}{arXiv:hep-lat/0305014}].



\end{thebibliography}
\end{document}